\documentclass[9pt,technote]{IEEEtran}
\usepackage{multicol}
\usepackage{amsmath,amssymb,amsthm,mathrsfs,amsfonts,amstext}
\usepackage{graphicx,subfigure}
\usepackage[comma,numbers,square,sort&compress]{natbib}
\usepackage{CJK}
\usepackage{tikz}
\usepackage{textcomp}
\usetikzlibrary{shapes,arrows,shadows}
\usepackage{verbatim}

\usepackage{amsmath,bm,times}

\tikzset{%
  block/.style    = {draw, thick, rectangle, minimum height = 2.5em,
    minimum width = 3em, fill=red!20, drop shadow},
  sum/.style      = {draw, circle, node distance = 2cm, fill=blue!20}, 
  input/.style    = {coordinate}, 
  output/.style   = {coordinate} 
}

\tikzstyle{sensor}=[draw, fill=blue!20, text width=4em,
    text centered, minimum height=2.5em,drop shadow]
\tikzstyle{ann} = [above, text width=5em, text centered]
\tikzstyle{wa} = [block, thick, rectangle, fill=red!20, minimum height = 3em,
    minimum width = 3em, drop shadow]
\tikzstyle{sc} = [sensor, text width=13em, fill=red!20,
    minimum height=10em, rounded corners, drop shadow]
\tikzstyle{nt}=[sensor, text width=6.5em, fill=red!20,
    minimum height=4.5em, rounded corners, drop shadow]
\def\ban{\begin{eqnarray*}}
\def\ean{\end{eqnarray*}}

\def\bna{\begin{eqnarray}}
\def\ena{\end{eqnarray}}

\newtheorem{theorem}{Theorem}

\newtheorem{lemma}{Lemma}
\newtheorem{remark}{Remark}

\begin{document}
\begin{CJK*}{GBK}{song}

\title{Coordination Over Multi-Agent Networks With Unmeasurable States and Finite-Level Quantization}

\author{Yang~Meng, ~\IEEEmembership{Student Member,~IEEE},
        Tao~Li,~\IEEEmembership{Senior Member,~IEEE}
        and~Ji-Feng~Zhang,~\IEEEmembership{Fellow,~IEEE}
\IEEEcompsocitemizethanks{\IEEEcompsocthanksitem  This work was partially presented at the 19th IFAC World Congress, Cape Town, South Africa, August, 2014. This work was supported by the National Natural Science Foundation of China under Grant  61522310, Shanghai Rising-Star Program under grant 15QA1402000, and the National Key Basic Research Program of China (973 Program) under grant
2014CB845301.  Please address all the correspondences to Tao Li: Phone: +86-21-56331183, Fax: +86-21-56331183, Email: sixumuzi@shu.edu.cn. \IEEEcompsocthanksitem Yang Meng is with the Key Laboratory of Systems and Control,
Academy of Mathematics and Systems Science, Chinese Academy of
Sciences, Beijing 100190, China. This work was accomplished when Yang Meng was a Research Assistant of the School of Mechatronic Engineering and Automation, Shanghai University.
\IEEEcompsocthanksitem Tao Li is with the Shanghai Key Laboratory of Power Station Automation Technology, School of Mechatronic  Engineering and Automation, Shanghai University, Shanghai 200072, China.
\IEEEcompsocthanksitem Ji-Feng Zhang is with the Key Laboratory of Systems and Control,
Academy of Mathematics and Systems Science, Chinese Academy of
Sciences, Beijing 100190, China. }}

\maketitle

\begin{abstract}
In this note, the coordination of linear discrete-time multi-agent systems over digital networks is
investigated with unmeasurable states in agents' dynamics.
The quantized-observer based communication protocols and Certainty
Equivalence principle based control protocols are proposed to characterize the
inter-agent communication and the cooperative control in an integrative framework.
By investigating the structural and asymptotic properties of the equations of stabilization and estimation errors, which are nonlinearly coupled
by the finite-level quantization scheme,
some necessary conditions and sufficient conditions are given for the
existence of such communication and control protocols to ensure the inter-agent state observation and cooperative stabilization. It is shown that these conditions come down to the simultaneous stabilizability and the detectability of the dynamics
of agents and the structure of the communication network.
\end{abstract}


\section{Introduction}
In recent years, the coordination of multi-agent systems
has attracted lots of attention by the systems and control community
due to its wide applications. 
For the coordination of multi-agent systems with digital
networks, the inter-agent communication, which aims at
obtaining neighbors' state information as precise as possible, is
usually the foundation of designing the cooperative control laws.
In real digital networks, communication channels only have finite
capacities and the communication between different agents is a
process which consists of encoding, information transmitting and
decoding. For this case, the instantaneously precise communication
is generally impossible and one may seek encoding-decoding
schemes to achieve asymptotically precise communication.

The most basic coordination of multi-agent systems is
distributed consensus or synchronization, which is also called
cooperative stabilization \cite{Qu (2009)}. Quantized consensus
and consensus with quantized communication can be dated back to
\cite{Kashyap Basar and Srikant (2007)} and \cite{Frasca Carli Fagnani and Zampieri (2009)} with the static quantization.
Carli {\it et.al.} \cite{Carli Bullo and Zampieri (2010)} proposed a dynamic encoding-decoding
scheme for distributed averaging. They proved that with
infinite-level logarithmic quantization, the closed-loop system can
achieve exact average-consensus asymptotically.
Li {\it et.al.} \cite{Li Fu Xie and Zhang (2011)} proposed a dynamic encoding-decoding
scheme with vanishing scaling function and finite-level uniform
quantizers. They proved that the exact average-consensus can be
achieved exponentially fast based on merely one-bit information
exchange per communication between neighbors. This algorithm was
then further generalized to the cases with directed and time-varying topologies (\cite{Li and Xie
(2011)}-\cite{Li Liu Wang and Yin (2014)}), the case with time delays (\cite{Liu Li and Xie (2011)}), the case with general linear agent dynamics with full measurable states (\cite{You and Xie (2011)}) and the case with second-order integrator dynamics with
partially measurable states (\cite{Li and Xie (2012)}). Recent works in this direction can be found in \cite{Olshevsky (2014)} for ternary information exchange, the continuous-time dynamics (\cite{Frasca (2012)})  and consensus over finite fields (\cite{Pasqualetti Borra and Bullo (2014)}).

All the above literature (\cite{Kashyap Basar and Srikant (2007)}-\cite{Li and Xie (2012)}) focused on designing
specific communication and control protocols and
analyzing the closed-loop performances for specific systems.
However, a fundamental problem of the coordination of multi-agent
systems over digital networks is for what kinds of dynamic
networks, there exist proper communication and control protocols
which can guarantee the objectives of the
inter-agent communication and cooperative control jointly. The
coordination of digital multi-agent networks consists of two
fundamental factors, one is the inter-agent state observation by
communication among agents, and the other one is the cooperative
control by each agent to achieve given coordination
objectives. The inter-agent state observation is the objective of
the inter-agent communication and is the basis of designing the cooperative
control laws. This is similar in spirit to that the state observation is
the basis of the feedback control design for single-agent systems
with unmeasurable states. It is of theoretical and practical
significance to characterize the inter-agent state observation and the
cooperative control of multi-agent systems in an integrative
framework. In this framework,  one needs to first give the conditions for the existence of
communication and control protocols to ensure both the communication and
control objectives. For the case with precise communication, the
consentability of linear multi-agent systems were studied. The
concept of consentability was first proposed by \cite{Zhang and Tian (2009)}-\cite{Ma and Zhang (2010)}.
It was shown that the controllability of agent dynamics and
the connectivity of the communication topology graph have a joint influence on the
consentability. You and Xie \cite{You and Xie (2011)} and Gu {\it et.al.} \cite{Gu Marinovici and Lewis (2012)} studied
the consentability of single-input linear discrete-time systems and
sufficient conditions were given with respect to (w. r. t.) relative
state feedback control protocols in \cite{You and Xie (2011)} and w. r. t.
filtered relative state feedback control protocols in \cite{Gu Marinovici and Lewis (2012)}, respectively.

In this note, motivated by \cite{Li and Xie (2012)}-\cite{You and Xie (2011)},
we consider the cooperatability of linear
discrete-time multi-agent systems with unmeasurable states and
finite communication data rate. We propose a class of communication
protocols based on quantized-observer type encoders and decoders and
a class of control protocols based on the relative state
feedback control law and the Certainty Equivalence principle.
The closed-loop dynamics of the cooperative stabilization and the state estimation errors are coupled by the nonlinearities generated by the finite-level quantization scheme. By investigating the structural and asymptotic properties of the overall closed-loop
equations, we give some necessary conditions and sufficient conditions for
achieving inter-agent state observation and cooperative
stabilization jointly w. r. t. the proposed classes of communication and control
protocols. It is shown that the cooperatability of multi-agent systems is related to the simultaneous
stabilizability and the detectability of the dynamics of agents and the structure of the
communication graph.

Different from \cite{You and Xie (2011)} for the case with fully measurable states, we consider
the case with unmeasurable states and the finite communication data rate. The
quantized-observer type encoding/decoding scheme proposed for
second-order integrators in \cite{Li and Xie (2012)} is generalized for the case with general linear dynamics.
Compared with \cite{You and Xie (2011)} and \cite{Gu Marinovici and Lewis (2012)} which focused on sufficient
conditions, we show that the simultaneous stabilizablility of $(A,\lambda_{i}(\mathcal{L})B),~i=2,\cdots,N$ and the detectability of $(A, C)$ are sufficient, and also necessary in some sense, for the cooperatability of the linear multi-agent systems over digital networks, where $A$, $B$ and $C$ are the system matrix, the input matrix and the output matrix, respectively, of each agent and $\lambda_{i}(\mathcal{L}),~i=2,\cdots,N$, are nonzero eigenvalues of the Laplacian matrix $\mathcal{L}$ of the communication graph.
We also show that the stabilizability of $(A, B)$(detectability of $(A,C)$) is
necessary for the cooperative stabilization (inter-agent state observation), regardless of
whether the inter-agent state observation (cooperative stabilization) is required.

The following notation will be used. Denote the column vectors or
matrices with all elements being $1$ and $0$ by $\textbf{1}$ and
$\textbf{0}$, respectively.  Denote the identity matrix with
dimension $n$ by $I_{n}$. Denote the sets of real
numbers, positive real numbers and conjugate numbers by $\mathbb{R}$, $\mathbb{R}^{+}$ and $\mathbb{C}$,
respectively, and $\mathbb{R}^{n}$ denotes the $n$-dimensional real
space. For any given vector $X\in\mathbb{R}^{n}$ or matrix
$X=[x_{ij}]\in\mathbb{R}^{n\times m}$, its transpose is denoted by
$X^{T}$, and its conjugate transpose is denoted by $X^{*}$. Denote
the Euclidean norm of $X$ by $\|X\|$ and the infinite norm of $X$ by $\|X\|_{\infty}$.
Denote the $k$th element of vector $X$ by $[X]_{k}$. Denote the spectral radius of square matrix $X$
by $\rho(X)$. Define
$\mathscr{B}^{n\times m}_{r}=\{X\in \mathbb{R}^{n\times m}|\|X\|<
r\}$ and $\mathscr{B}^{n}_{r}=\{X\in \mathbb{R}^{n}|\|X\|_{\infty}<
r\}$, $r\in\mathbb{R}^{+}\bigcup\{+\infty\}$.
The Kronecker product is denoted by $\otimes$.

\section{Problem Formulation}\label{II}
The dynamics of each agent is
given by
\begin{equation}\label{1}
    \left\{
\begin{aligned}
& x_{i}(t+1)=Ax_{i}(t)+Bu_{i}(t),\\
& y_{i}(t)=Cx_{i}(t),
\end{aligned}\right.\qquad t=0,1,\cdots,
\end{equation}
where $A\in \mathbb{R}^{n\times n}$, $B\in\mathbb{R}^{n\times m}$ and
$C\in \mathbb{R}^{p\times n}$. Here, $x_{i}(t)$, $y_{i}(t)$ and $u_{i}(t)$
are the state, the output and the control input of agent $i$.
The overall communication structure of the network is represented by
a directed graph $\mathscr{G}=\{V,\mathscr{E},\mathscr{A}\}$, where
$V=\{1,\cdots,N\}$ is the node set and each node represents an
agent; $\mathscr{E}$ denotes the edge set and there is an edge
$(j,i)\in \mathscr{E}$ if and only if there is a communication
channel from $j$ to $i$, then, agent $i$ is called the receiver and
agent $j$ is called the sender, or $i$'s neighbor. The set of
agent $i$'s neighbors is denoted by $N_{i}=\{j\in
V|(j,i)\in\mathscr{E}\}$.
We denote $\mathscr{A}=[a_{ij}]\in \mathbb{R}^{N\times N}$ as the
weighted adjacent matrix of $\mathscr{G}$, $a_{ij}>0$ if and only if
$j\in N_{i}$. Here we assume $a_{ii}=0$. Denote
$deg_{i}=\sum_{j=1}^{N}a_{ij}$ as the in-degree of node $i$ and
$\mathscr{D}=diag(deg_{1},\cdots,deg_{N})$ is called the degree
matrix of $\mathscr{G}$. The Laplacian matrix $\mathcal{L}$ of
$\mathscr{G}$ is defined as $\mathcal{L}=\mathscr{D}-\mathscr{A}$,
and its eigenvalues in an ascending order of real parts are denoted
by $\lambda_{1}(\mathcal{L})=0$, $\lambda_{i}(\mathcal{L})$, $i=2,
\cdots,N$. The agent dynamics (\ref{1}) together with the communication topology graph $\mathscr{G}$ is called a dynamic network\footnote{The concept of dynamic network of agents without output equations was defined in \cite{Olfati-Saber and Murray (2004)}.} and is denoted by $(A, B, C, \mathscr{G})$.

For real digital networks, only finite bits of data can be transmitted
at each time step, therefore, each agent needs to first quantize and
encode its output into finite symbols before transmitting them.
Each pair of adjacent agents uses an encoding-decoding scheme
to exchange information: For each digital communication channel
$(j,i)$, there is an encoder/decoder pair, denoted by $H_{ji}=(\Theta_{j},\Psi_{ji})$,
associate with it. Here, $\Theta_{j}$ denotes the encoder maintained by agent $j$ and $\Psi_{ji}$ denotes the
decoder maintained by agent $i$.
For the dynamic network $(A, B, C, \mathscr{G})$, the set $\{H_{ji},i=1,\cdots, N, j\in
N_{i}|H_{ji}=(\Theta_{j},\Psi_{ji})\}$ of encoder-decoder pairs over the whole network
is called a communication
protocol, and the collection of such communication protocols
is denoted by the communication protocol set
$\mathscr{H}$.

In this note, we propose the following communication protocol set:
\begin{equation}\label{rongxutongxinxieyiji}
\begin{aligned}
\mathscr{H}(\varrho, L_{G})&=\Big\{H(\gamma, \alpha, \alpha_{u}, L, L_{u}, G),
\gamma\in(0,\varrho), \alpha\in(0,1],\\
&~~~~~~~~\alpha_{u}\in(0,1],
 L\in\mathbb{N}, L_u\in\mathbb{N}, G\in\mathscr{B}^{n\times p}_{L_{G}}\Big\},
\end{aligned}
\end{equation}
where
$H(\gamma, \alpha, \alpha_{u}, L, L_u, G)=\{H_{ji}, i=1, \cdots, N, j\in N_{i}|H_{ji}=(\Theta_{j},\Psi_{ji}), \}$.
Here the constants $L_{G}\in\mathbb{R}^{+}\bigcup\{+\infty\}$, $\varrho\in(0,1]$ are given parameters of the
communication protocol set, while $\gamma$,  $\alpha$,
$\alpha_{u}$, $L$, $L_u$ and $G$ are parameters of a specific communication
protocol. For each digital channel $(j,i)$, the encoder is given by
\begin{equation}\label{bianma}
\Theta_{j}=\left\{
\begin{aligned}
& \hat{x}_{j}(0)=\hat{x}_{j0},\ \hat{u}_{j}(0)=\hat{u}_{j0},\\
& s_{j}(t)=Q_{\alpha, L}\left(\frac{y_{j}(t-1)-C\hat{x}_{j}(t-1)}{\gamma^{t-1}}\right),\\
& \hat{x}_{j}(t)=A\hat{x}_{j}(t-1)+\gamma^{t-1}Gs_{j}(t)+B\hat{u}_{j}(t-1),\\
& \hat{u}_{j}(t)=\hat{u}_{j}(t-1)+\gamma^{t-1}s_{u,j}(t),\\
& s_{u,j}(t)=Q_{\alpha_u, L_u}\left(\frac{u_{j}(t)-\hat{u}_{j}(t-1)}{\gamma^{t-1}}\right).
\end{aligned}
\right.
\end{equation}
and the decoder is given by
\begin{equation}\label{jiema}
\Psi_{ji}=\left\{
\begin{aligned}
& \hat{x}_{ji}(0)=\hat{x}_{j0},\ \hat{u}_{ji}(0)=\hat{u}_{j0},\\
& \hat{x}_{ji}(t)=A\hat{x}_{ji}(t-1)+\gamma^{t-1}Gs_{j}(t)+B\hat{u}_{ji}(t-1),\\
& \hat{u}_{ji}(t)=\hat{u}_{ji}(t-1)+\gamma^{t-1}s_{u,j}(t),
\end{aligned}\right.
\end{equation}
where $Q_{p, M}(\cdot)$ with $p\in(0,1]$ and $M=1,2,...$ is a finite-level uniform
quantizer. For vector inputs, the definition is applied to each component.
\begin{equation*}
    Q_{p, M}(y)=\left\{
\begin{aligned}
& ip,\qquad ip-\frac{p}{2}\leq y<ip+\frac{p}{2},\ i=0,1,..., M-1\\
& Mp,\qquad y\geq Mp-\frac{p}{2},\\
& -Q_{p,M}(-y),\qquad y<-\frac{p}{2}.
\end{aligned}\right.
\end{equation*}
At each time step $t$, agent $j$ generates the symbolic data $s_{j}(t)$ and $s_{u,j}(t)$ by
the encoder $\Theta_{j}$ and sends them to agent $i$ through
the channel $(j,i)$. After $s_{j}(t)$, $s_{u,j}(t)$ are received, by
the decoder $\Psi_{ji}$, agent $i$ calculates $\hat{x}_{ji}(t)$ as
an estimate of $x_{j}(t)$. Denote
$E_{ji}(t)=x_{j}(t)-\hat{x}_{ji}(t)$ as the state estimation error. From (\ref{bianma}) and (\ref{jiema}),
we have $E_{ji}(t)=x_{j}(t)-\hat{x}_{j}(t)$, and is denoted by $E_{j}(t)$ for
short. Here, we say that the dynamic network achieves inter-agent state observation if
\begin{equation*}
\lim_{t\rightarrow\infty}(x_{j}(t)-\hat{x}_{ji}(t))=\textbf{0},\ i=1, \cdots,
N,~ j\in N_{i}.
\end{equation*}

For the case with precise communication, Olfati-Saber and Murray
\cite{Olfati-Saber and Murray (2004)} proposed a class of relative state
feedback control protocols :
\begin{equation}\label{jingquekongzhilv}
    u_{i}(t)=K\sum_{j=1}^{N}a_{ij}(x_{j}(t)-x_{i}(t)),\
    i=1, \cdots, N.
\end{equation}
Based on (\ref{jingquekongzhilv}) and the Certainty Equivalence
principle, we propose the following control protocol set: $\mathscr{U}(L_{K})=\{U(K),K\in\mathscr{B}^{m\times n}_{L_{K}}\}$, where
\begin{equation}
\label{rongxukongzhi}
\begin{aligned}
U(K)&=\Big\{u_{i}(t),t=0, 1,\cdots,
i=1,..., N|\\
&~~~~~u_{i}(t)=K\sum_{j\in N_i}a_{ij}(\hat{x}_{ji}(t)-\hat{x}_{i}(t)) \Big\}.
\end{aligned}
\end{equation}
The constant
$L_{K}\in\mathbb{R}^{+}\bigcup\{+\infty\}$ is the given parameter of
the control protocol set and the gain matrix $K$ is the
parameter of a control protocol to be designed.

We say that the dynamic network $(A, B, C, \mathscr{G})$ is
locally cooperatable
if for any given positive constants $C_{1}$, $C_{2}$, $C_{3}$, there exist communication and control protocols
$H\in\mathscr{H}$ and $U\in\mathscr{U}$, such that for any $x_{i}(0)\in\mathscr{B}^{n}_{C_{1}}$,
$\hat{x}_{i0}\in\mathscr{B}^{n}_{C_{2}}$, and
$\hat{u}_{i0}\in\mathscr{B}^{m}_{C_{3}}$, $i\in
1,\cdots, N$,  the closed-loop system achieves
inter-agent state observation and cooperative stabilization.
that is,
\ban
 &&(a) \ \lim_{t\rightarrow\infty}E_{j}(t)=\textbf{0},\ j=1, \cdots, N.\cr
 &&(b) \ \lim_{t\rightarrow\infty}(x_{i}(t)-x_{j}(t))=\textbf{0},\ i,j=1, \cdots, N.
\ean
The dynamic network  is called globally cooperatable
 if there exist communication and control protocols $H\in\mathscr{H}$
and $U\in\mathscr{U}$, such that for any given initial
condition, the closed-loop system achieves inter-agent state observation and cooperative
stabilization.

\vskip 0.2cm

\begin{remark}
{\rm Different from \cite{You and Xie (2011)} and \cite{Gu Marinovici and Lewis (2012)}, we consider the cooperatability of linear multi-agent
systems with unmeasurable states and finite data rate. A
quantized-observer based encoding-decoding scheme is proposed to
estimate neighbors' states while decoding. From
(\ref{jiema}), the decoder has a similar structure as the
Luenberger observer. For the case with precise communication, the
quantizers degenerate to identical functions and the decoders
degenerate to the Luenberger observers. }
\end{remark}

\vskip 0.2cm

\begin{remark}
{\rm
(i)
Our quantized observer is based on the quantized innovation of $y_{i}(t)$ but not $y_{i}(t)$ itself.
This type of observer is also called differential pulse code modulation (DPCM) in the communication community, which can save the bandwidth
of the communication channel significantly \cite{Li Fu Xie and Zhang (2011)}, \cite{Li and Xie (2012)}. (ii) In a single-agent system, the controller and observer are usually
located on the same side, which means the exact value of the control input can be used to design the observer directly. However, for
the inter-agent state observation of multi-agent systems, the observers for neighbors' states are located faraway from the neighbors' controllers,
which means the exact values of neighbors' control inputs are not available. Therefore, the estimations of neighbors' control inputs are added
into our encoding-decoding schemes.
(iii) For second-order integrator agents, the special dynamic structure makes it feasible to reconstruct neighbors' control inputs by differencing the delayed positions and velocities without explicitly estimate neighbors' control inputs (\cite{Li and Xie (2012)})
However, for general linear dynamics,  the method in \cite{Li and Xie (2012)} can not be used here. Here, we propose the Luenberger form decoders (\ref{jiema}) with explicit estimations of neighbors' control inputs.}
\end{remark}

\vskip 0.2cm

\begin{remark}
{\rm Here, as a preliminary research, the definition of cooperatability focus on the ability of multi-agent systems to achieve inter-agent state observation and cooperative stabilization. The cooperative stabilization (synchronization) is the most basic cooperation of multi-agent systems and forms the foundation of many other kinds of cooperative controls, such as formation and distributed tracking. The concept of cooperatability can be further expanded for more general coordination behaviors. One may wonder that to achieve synchronization, why we do not use the decentralized state feedback control law $u_{i}(t)=-Kx_{i}(t)$ for each agent, then all agents' states will go to zero without any inter-agent communication. We do not use the decentralized state feedback control law but (\ref{rongxukongzhi}) mainly for two points. (i) The decentralized state feedback control law leads to the trivial case, i. e. all agents' states will go to zero. Here, the closed-loop system can achieve more general behavior. One may see that all agents' states will approach the weighted average initial values multiplied by the exponent of the system matrix under the control protocol (\ref{rongxukongzhi}) (see also Remark \ref{zuizhongguiji}). This gives more flexibility to achieve complex coordination behavior by adjusting control and system parameters.
(ii) The control protocol (\ref{rongxukongzhi}) is more flexible than the decentralized state feedback control law. One may further extend it for the formation control based on relative state vectors (\cite{LWCV2005}):
\bna
\label{rongxukongzhiformation}
u_{i}(t)&=&K\sum_{j\in N_i}a_{ij}(\hat{x}_{ji}(t)-\hat{x}_{i}(t)-b_{ij}),
i=1,..., N,
\ena
or the distributed tracking problem:
\bna
\label{rongxukongzhitracking}
u_{i}(t)&=&K_{1}\sum_{j\in N_i}a_{ij}(\hat{x}_{ji}(t)-\hat{x}_{i}(t))\cr
&&+K_2b_{i0}(\hat{x}_{0i}(t)-\hat{x}_{i}(t)), i=1,..., N.
\ena
}
\end{remark}

\section{Main Results}\label{III}

In this section we give some necessary conditions and
sufficient conditions which ensure $(A,B,C,\mathscr{G})$  to be
cooperatable. The following assumptions will be used.

\vskip 0.2cm

\textbf{A1)} There exists $K\in \mathbb{R}^{m\times n}$ such that
the eigenvalues of $A-\lambda_{i}(\mathcal{L})BK$, $i=2, \cdots, N$
are all inside the open unit disk of the complex plane.

\vskip 0.2cm

\textbf{A2)} $(A,C)$ is detectable.

\vskip 0.2cm

Denote $\Delta_{j}(t-1)=\frac{y_{j}(t-1)-C\hat{x}_{j}(t-1)}{\gamma^{t-1}}-s_{j}(t)
$ and $\Delta_{u,j}(t-1)=\frac{u_{j}(t)-\hat{u}_{j}(t-1)}{\gamma^{t-1}}-s_{u,j}(t)$
as the quantization errors of $Q_{\alpha,L}(\cdot)$ and $Q_{\alpha_{u},L_{u}}(\cdot)$,
respectively. Denote $\Delta(t)=(\Delta_{1}^{T}(t),\cdots, \Delta_{N}^{T}(t))^{T}$,
$\Delta_{u}(t)=(\Delta_{u,1}^{T}(t),\cdots,\Delta_{u,N}^{T}(t))^{T}$. Denote
$X(t)=(x^{T}_{1}(t),\cdots,x^{T}_{N}(t))^{T}$, $\hat{X}(t)=
(\hat{x}^{T}_{1}(t),\cdots,\hat{x}^{T}_{N}(t))^{T}$,
$U(t)=(u^{T}_{1}(t) ,\cdots,u^{T}_{N}(t))^{T}$, $\hat{U}(t)=(\hat{u}^{T}_{1}(t),\cdots,$ $\hat{u}^{T}_{N}(t))^{T}$.
Denote $E(t)=X(t)-\hat{X}(t)$, $H(t)=U(t)-\hat{U}(t)$, $\bar{X}(t)=(\frac{1}{\mathbf{1}^{T}\pi}\mathbf{1}\pi^{T}\otimes I_{n})X(t)$,
$\delta(t)=X(t)-\bar{X}(t)$, where
$\pi^{T}$ is the nonnegative left eigenvector w. r. t. the eigenvalue $0$ of $\mathcal{L}$ and it can be verified that
$\pi^{T}$ has at least one nonzero element. Here, $\delta(t)$ is called the cooperative stabilization error.
Denote the lower triangular Jordan
canonical of $\mathcal{L}$ by $diag(0,J_{2},\cdots,J_{N})$ where
$J_{i}$ is the Jordan chain with respect to
$\lambda_{i}(\mathcal{L})$. We know that there is $\Phi\in\mathbb{R}^{N\times N}$, consisting of the left eigenvectors and generalized
left eigenvectors of $\mathcal{L}$, such that
$\Phi\mathcal{L}\Phi^{-1}=diag(0,J_{2},\cdots,J_{N})$. Let
$\Phi=(\pi,\phi_{2},\cdots,\phi_{N})^{T}$. Denote
$\bar{J}(K)=I_{N-1}\otimes A-diag(J_{2},\cdots,J_{N})\otimes BK$,
$J(G)=diag(A-GC,\cdots,A-GC)_{nN\times
 nN}$.

From $(\ref{1})$, $(\ref{bianma})$, $(\ref{jiema})$ and $(\ref{rongxukongzhi})$, we have
\begin{equation}\label{5}
X(t+1)=(I_{N}\otimes A)X(t)-(\mathcal{L}\otimes BK)\hat{X}(t).
\end{equation}
and
\begin{equation}\label{e t+1}
\begin{aligned}
E(t+1)&=(I_{N}\otimes(A-GC))E(t)+(I_{N}\otimes B)H(t)\\
&~\gamma^{t}(I_{N}\otimes G)\Delta(t).
\end{aligned}
\end{equation}
Note that $\pi^{T}\mathcal{L}=0$, from $(\ref{5})$ it is known that $\bar{X}(t+1)=(I_{N}\otimes A)\bar{X}(t)$.
 Thus, from $(\ref{5})$, the definition of $E(t)$, $\delta(t)$ and noting that $\mathcal{L}\mathbf{1}=0$, we have
\begin{equation}\label{delta t+1}
    \begin{aligned}
    \delta(t+1)=(I_{N}\otimes A-\mathcal{L}\otimes BK)\delta(t)+(\mathcal{L}\otimes
    BK)E(t).
   \end{aligned}
\end{equation}
Denote $F(t)=\frac{U(t+1)-\hat{U}(t)}{\gamma^{t}}$. By $(\ref{bianma})$ and the definition of $H(t)$, we have
\begin{equation}\label{h t+1}
    \begin{aligned}
    H(t+1)&=U(t+1)-\hat{U}(t)-\gamma^{t}Q_{\alpha_{u},L_{u}}\left(\frac{U(t+1)-\hat{U}(t)}{\gamma^{t}}\right)\\
    &=\gamma^{t}(F(t)-Q_{\alpha_{u},L_{u}}(F(t)))=\gamma^{t}\Delta_{u}(t).
    \end{aligned}
    \end{equation}
From $(\ref{rongxukongzhi})$, $(\ref{e t+1})$ and $(\ref{delta t+1})$ we can
see that
\begin{equation}
\begin{aligned}
F(t)&=(\mathcal{L}\otimes K-\mathcal{L}\otimes KA+\mathcal{L}^{2}\otimes
    KBK)\frac{\delta(t)}{\gamma^{t}}\\
&~+(\mathcal{L}\otimes K(A-GC)-\mathcal{L}^{2}\otimes KBK-\mathcal{L}\otimes
    K)\frac{E(t)}{\gamma^{t}}\\
    &~+(I_{mN}+\mathcal{L}\otimes
    KB)\frac{H(t)}{\gamma^{t}}+(\mathcal{L}\otimes
    KG)\Delta(t)
\end{aligned}
\end{equation}
Thus, we have the following equations:
\begin{equation}\label{FULU}
\left\{
\begin{aligned}
    &E(t+1)=(I_{N}\otimes(A-GC))E(t)+(I_{N}\otimes B)H(t)\\
    &~~~~~~~~~~~~+\gamma^{t}(I_{N}\otimes
    G)\Delta(t),\\
    &\delta(t+1)=(I_{N}\otimes A-\mathcal{L}\otimes BK)\delta(t)+(\mathcal{L}\otimes
    BK)E(t),\\
    &H(t+1)=\gamma^{t}(F(t)-Q_{\alpha_{u},L_{u}}(F(t))),\\
    &F(t)=(\mathcal{L}\otimes K-\mathcal{L}\otimes KA+\mathcal{L}^{2}\otimes
    KBK)\frac{\delta(t)}{\gamma^{t}}\\
    &~~~~~~~+(\mathcal{L}\otimes K(A-GC)-\mathcal{L}^{2}\otimes KBK-\mathcal{L}\otimes
    K)\frac{E(t)}{\gamma^{t}}\\
    &~~~~~~~+(I_{mN}+\mathcal{L}\otimes
    KB)\frac{H(t)}{\gamma^{t}}+(\mathcal{L}\otimes
    KG)\Delta(t).
\end{aligned}\right.
\end{equation}

We have the following theorems. The proofs of Theorems 1, 2, 4 and 5 are put in Appendix.

\vskip 0.2cm

\begin{theorem}\label{youxiangyinli}
For the dynamic network $(A, B, C, \mathscr{G})$, $L_{G}=+\infty$, $\varrho=1$ and
$L_{K}=+\infty$,  suppose that Assumptions \textbf{A1)} and \textbf{A2)}
hold. Then, for any given positive constants $C_{x}$, $C_{\hat{x}}$
and $C_{\hat{u}}$, there exist a communication protocol $H(\gamma,
\alpha, \alpha_{u}, L, L_u, G)\in\mathscr{H}(\varrho, L_{G})$ and a
control protocol $U(K)\in\mathscr{U}(L_{K})$ such that for any
$X(0)\in \mathscr{B}^{nN}_{C_{x}}$, $\hat{X}(0)\in
\mathscr{B}^{nN}_{C_{\hat{x}}}$ and
$\hat{U}(0)\in\mathscr{B}^{mN}_{C_{\hat{u}}}$, the dynamic network
$(A,B,C,\mathscr{G})$ achieves inter-agent state observation and
cooperative stabilization under $H$ and $U$, and there exist
positive constants $W$ and $W_{u}$ independent of $\gamma$,
$\alpha$, $\alpha_{u}$, $L$, $L_u$, $G$ and $K$, such that
$\sup_{t\geq0}\max_{1\leq j\leq
N}\|\Delta_{j}(t)\|_{\infty}\leq W$ and
$\sup_{t\geq0}\max_{1\leq j\leq
N}\|\Delta_{u,j}(t)\|_{\infty}\leq W_{u}$.
\end{theorem}

\vskip 0.2cm

\begin{remark}\label{zuizhongguiji}
{\rm
It can be verified that
$\bar{X}(t+1)=(I_{N}\otimes A)\bar{X}(t), ~t=0,1,2,\cdots$.
Since $\lim_{t\rightarrow\infty}\delta(t)=0$, we have
\begin{equation*}
\begin{aligned}
\lim_{t\rightarrow\infty}\left[x_{i}(t)-A^{t}\Big(\frac{\sum_{i=1}^{N}\pi_{i}x_{i}(0)}{\sum_{i=1}^{N}\pi_{i}}\Big)\right]=0, ~i=1,\cdots,N.
\end{aligned}
\end{equation*}
So all agents' states will finally approach the trajectory $A^{t}\Big(\frac{\sum_{i=1}^{N}\pi_{i}x_{i}(0)}{\sum_{i=1}^{N}\pi_{i}}\Big)$.
If the control protocol (\ref{rongxukongzhi}) is replaced by
\bna
\label{rongxukongzhiabsolutestate}
u_{i}(t)=K_{1}x_{i}(t)+K_2\sum_{j\in N_i}a_{ij}(\hat{x}_{ji}(t)-\hat{x}_{i}(t)),
\cr
t=0, 1,\cdots,
i=1,..., N.
\ena
which combines the decentralized state feedback and the distributed quantized relative output feedback, then the closed-loop states approach $(A+BK_1)^{t}\Big(\frac{\sum_{i=1}^{N}\pi_{i}x_{i}(0)}{\sum_{i=1}^{N}\pi_{i}}\Big)$. For this kind of control  protocols, one may choose $K_1$ to achieve more complex coordination behavior.}
\end{remark}

\vskip 0.2cm

\begin{remark}
{\rm Intuitively, Assumption \textbf{A1)} contains the requirement on the agent
dynamics $(A, B)$ and the communication topology graph
$\mathscr{G}$. If $\rho(A)<1$, cooperative stabilization can be achieved by taking $K=\textbf{0}$ (leading to a trivial case),
which makes $A-\lambda_{i}(\mathcal{L})BK=A$, $i=2, \cdots, N$ all
stable even $\mathscr{G}$ has no spanning tree ($\lambda_{2}(\mathcal{L})=0$). If $\rho(A)\geq1$, then Assumption \textbf{A1)} requires that
$\lambda_{2}(\mathcal{L})\neq0$, which implies that $\mathscr{G}$
contains a spanning tree \cite{Ren and Beard (2008)}.

For single input discrete-time systems, \cite{You and Xie (2011)} gave a necessary and sufficient condition to ensure \textbf{A1)} if all of $A$'s eigenvalues are on or outside the unit circle of the complex plane, which was a intuitional explanation of \textbf{A1)}. In fact, for single input agents, a sufficient condition to ensure \textbf{A1)} can be given:

\vskip 0.2cm

\textbf{A1$^{'}$)} $(A,B)$ is stabilizable and
$$\prod_{j}|\lambda_{j}^{u}(A)|< \frac{1}{\inf_{\omega\in
\mathbb{R}}\max_{j\in\{2,\cdots,N\}}|1-\omega\lambda_{j}(\mathcal{L})|}.$$

\vskip 0.2cm

Here, $\lambda_{j}^{u}(A)$, $1\leq j\leq n$
denote the unstable eigenvalues of $A$. If $\rho(A)<1$, then $\prod_{j}|\lambda_{j}^{u}(A)|$ is defined as $0$. What's more, if the communication topology graph is undirected, it was shown in \cite{You and Xie (2011)} that $\frac{1}{\inf_{\omega\in
\mathbb{R}}\max_{j\in\{2,\cdots,N\}}|1-\omega\lambda_{j}(\mathcal{L})|}=\frac{1+\lambda_{2}/\lambda_{N}}{1-\lambda_{2}/\lambda_{N}}$ and thus the eigenvalue-ratio $\lambda_{2}/\lambda_{N}$ plays an important part in the cooperatability of linear multi-agent systems.}
\end{remark}

\vskip 0.2cm

The following theorem shows that Assumption \textbf{A1$^{'}$)} implies \textbf{A1)}.


\begin{theorem} 
\label{youkeyoutuilun}
For single input agents, if Assumption \textbf{A1$^{'}$)} holds, then Assumption \textbf{A1)} holds.
\end{theorem}

\vskip 0.2cm

Theorem \ref{youxiangyinli} shows that Assumptions \textbf{A1)} and \textbf{A2)} are
sufficient conditions for the cooperatability of
$(A,B,C,\mathscr{G})$. Furthermore, we find that they are also
necessary conditions if $\varrho<1$.

\vskip 0.2cm

\begin{theorem}\label{dingli}
For $(A,B,C,\mathscr{G})$ and $L_{G}>0$, $L_{K}>0$ and
$\varrho\in(0,1)$, suppose that for any given positive constants $C_{x}$,
$C_{\hat{x}}$ and $C_{\hat{u}}$, there exist a communication
protocol $H(\gamma, \alpha, \alpha_{u}, L, L_u,
G)\in\mathscr{H}(\varrho, L_{G})$ and a control protocol
$U(K)\in\mathscr{U}(L_{K})$, such that for any $X(0)\in
\mathscr{B}^{nN}_{C_{x}}$, $\hat{X}(0)\in
\mathscr{B}^{nN}_{C_{\hat{x}}}$ and
$\hat{U}(0)\in\mathscr{B}^{mN}_{C_{\hat{u}}}$, the closed-loop
system achieves inter-agent state observation and cooperative
stabilization under $H$ and $U$, and the quantization errors satisfy
$\sup_{t\geq0}\max_{1\leq j\leq
N}\|\Delta_{j}(t)\|_{\infty}\leq W$ and
$\sup_{t\geq0}\max_{1\leq j\leq
N}\|\Delta_{u,j}(t)\|_{\infty}$ $\leq W_{u}$, where $W$ and $W_{u}$ are
positive constants independent of $\gamma$, $\alpha$, $\alpha_{u}$,
$L$, $L_u$, $G$ and $K$. Then Assumptions \textbf{A1)} and \textbf{A2)} hold.
\end{theorem}
\textbf{Proof}:
We will use reduction to absurdity. Suppose that for any positive
$C_{x},C_{\hat{x}},C_{\hat{u}}$, there exist a communication
protocol $H(\gamma, \alpha, \alpha_{u}, L, L_u,
G)\in\mathscr{H}(\varrho, L_{G})$ and a control protocol
$U(K)\in\mathscr{U}(L_{K})$ such that under these protocols, for any
$X(0)\in \mathscr{B}^{nN}_{C_{x}}$, $\hat{X}(0)\in
\mathscr{B}^{nN}_{C_{\hat{x}}}$ and $\hat{U}(0)\in\mathscr{B}^{mN}_{C_{\hat{u}}}$, the closed-loop system
satisfies $\lim_{t\rightarrow\infty}E(t)=0$, $\lim_{t\rightarrow\infty}\delta(t)=0$, $\sup_{t\geq0}\max_{1\leq
j\leq N}\|\Delta_{j}(t)\|_{\infty}$ $\leq W$ and
$\sup_{t\geq0}\max_{1\leq j\leq N}\|\Delta_{u,j}(t)\|_{\infty}\leq W_{u}$, however, \textbf{A1)} or \textbf{A2)} would not hold.
Select a constant $a$ satisfying
\begin{equation}\label{aaa}
a>\frac{4W_{u}\|B\|\sqrt{mN}}{1-\varrho}+\frac{4L_{G}W\sqrt{nN}}{1-\varrho}.
\end{equation}
Take $C_{x}>\sqrt{n(2N-1)}a\|\Phi^{-1}\|$, $C_{\hat{x}}>\sqrt{nN}C_{x}+a\sqrt{nN}$ and
$C_{\hat{u}}>\sup_{K\in \mathscr{B}_{L_{K}}}\| \mathcal{L}\otimes
K\|C_{\hat{x}}\sqrt{nN}$. Now we prove that if \textbf{A1)} or \textbf{A2)} would not hold, then for such
$C_{x}$, $C_{\hat{x}}$ and $C_{\hat{u}}$, there exist $X(0)\in\mathscr{B}^{nN}_{C_{x}}$, $\hat{X}(0)\in\mathscr{B}^{nN}_{C_{\hat{x}}}$
and $\hat{U}(0)\in\mathscr{B}^{mN}_{C_{\hat{u}}}$ such that under any communication protocol in (\ref{rongxutongxinxieyiji}) and control protocol in (\ref{rongxukongzhi}), the dynamic
network can not achieve inter-agent state observation and cooperative
stabilization jointly, which
leads to the contradiction.

Denote $\tilde{\delta}(t)=(\Phi\otimes I_{n})\delta(t)$. Denote $\bar{\Phi}=(\phi_{2},\cdots,\phi_{N})^{T}$, and denote $\tilde{\delta}_{2}(t)=(\bar{\Phi}\otimes I_{n})\delta(t)$. From
(\ref{FULU}), it follows that
\begin{equation}\label{E,DELTA2}
\begin{aligned}
\left(\begin{array}{c}
        E(t+1) \\
        \tilde{\delta}_{2}(t+1)
      \end{array}
\right)&=A(K,G)\left(\begin{array}{c}
               E(t) \\
               \tilde{\delta}_{2}(t)
             \end{array}
\right)+\left(\begin{array}{c}
                I_{nN} \\
                \mathbf{0}
              \end{array}
\right)(I_{N}\otimes B)\\
&~\cdot H(t)+\left(\begin{array}{c}
                                    I_{nN} \\
                                    \mathbf{0}
                                  \end{array}
\right)(I_{N}\otimes G)\gamma^{t}\Delta(t),
\end{aligned}
 \end{equation}
where $A(K,G)=\left(\begin{array}{cc}
                J(G) & \mathbf{0} \\
                (\bar{\Phi}\otimes I_{n})(\mathcal{L}\otimes BK) &
                \bar{J}(K)
              \end{array}
\right)$. Since \textbf{A1)} and \textbf{A2)} would not hold simultaneously, we have
$\rho(A(K,G))\geq1$ under any communication protocol in (\ref{rongxutongxinxieyiji}) and control protocol in (\ref{rongxukongzhi}). Transform $A(K,G)$ to its Schur canonical, that is,
select a unitary matrix $P$ ($P^{*}=P^{-1}$) such that
\begin{equation*}
\begin{aligned}
   &P^{*}A(K,G)P\\
   &~~=\left(\begin{array}{ccc}
                        \lambda_{1}(A(K,G)) &  & \mathbf{0 }\\
                         \times & \ddots &  \\
                        \times & \times  & \lambda_{(2N-1)n}(A(K,G))
                      \end{array}
   \right).
   \end{aligned}
\end{equation*}
Here, $\lambda_{1}(A(K,G))$, $\cdots$,
$\lambda_{(2N-1)n}(A(K,G))$ are eigenvalues of $A(K,G)$ with $|\lambda_{1}(A(K,G))|=\rho(A(K,G))$, and $\times$ represents the elements below the diagonal of the
Schur canonical.

Denote $Z(t)=P^{*}[E^T(t),\tilde{\delta}_{2}^T(t)]^T$. From $(\ref{E,DELTA2})$ we
know that
\begin{equation}
\begin{aligned}
\label{z1}
&[Z(t+1)]_{1}\\
&~=\lambda^{t+1}_{1}(A(K,G))[Z(0)]_{1}\\
&~~+\sum_{i=1}^{t}\lambda_{1}^{t-i}(A(K,G))\big[P^{*}[I_{nN},\mathbf{0}^T]^T
             (I_{N}\otimes
B)H(i)\big]_{1}\\
&~~+\sum_{i=0}^{t}\lambda_{1}^{t-i}(A(K,G))\gamma^{i}\big[P^{*}
                                    [I_{nN},\mathbf{0}^T]^T
                                 (I_{N}\otimes G)\Delta(i)\big]_{1}\\
&~~+\lambda_{1}^{t}(A(K,G))\big[P^{*}[I_{nN},\mathbf{0}^T]^T(I_{N}\otimes B)H(0)\big]_{1}.
\end{aligned}
\end{equation}
Let $P=[P_{1}^T, P_{2}^T]^T$ with $P_{1}\in\mathbb{R}^{nN\times n(2N-1)}$ and $P_{2}\in \mathbb{R}^{n(N-1)\times n(2N-1)}$. Take $X(0)=(\Phi^{-1}\otimes I_{n})[\mathbf{0}^T,\mathbf{a}^TP_{2}^T]^T$ where $\mathbf{a}=a\mathbf{1}\in\mathbb{R}^{n(2N-1)}$ and $\mathbf{0}\in\mathbb{R}^{n}$, then $\|X(0)\|_{\infty}\leq
\sqrt{n(2N-1)}a\|\Phi^{-1}\| \| P_{2}\|$. Note that $\|
P_{2}\|\leq\| P\|=1$, we have $\|X(0)\|_{\infty}\leq
\sqrt{n(2N-1)}a\|\Phi^{-1}\|<C_{x}$, implying
$X(0)\in\mathscr{B}^{nN}_{C_{x}}$. Take $\hat{X}(0)=X(0)-P_{1}\mathbf{a}$
and $\hat{U}(0)=-(\mathcal{L}\otimes K)\hat{X}(0)$. Similarly, one can see that $\hat{X}(0)\in\mathscr{B}^{nN}_{C_{\hat{x}}}$
and $ \hat{U}(0)\in\mathscr{B}^{mN}_{C_{\hat{u}}}$. By the definition of $\delta(t)$ and some direct calculation, we have
$\tilde{\delta}(0)=[\mathbf{0}^T,\mathbf{a}^TP_{2}^T]^T$, and $\tilde{\delta}_{2}(0)=P_{2}\mathbf{a}$. By the definition of $E(t)$ and
    $H(t)$, we know that $E(0)=X(0)-\hat{X}(0)=X(0)-(X(0)-P_{1}\mathbf{a})=P_{1}\mathbf{a}$,
    and $H(0)=U(0)-\hat{U}(0)=-(\mathcal{L}\otimes K)\hat{X}(0)+(\mathcal{L}\otimes K)\hat{X}(0)=\mathbf{0}$. Since
    $Z(0)=P^{*}[
                 E(0)^{T},
                 \tilde{\delta}_{2}^{T}(0)
]^{T}
    $, we have $Z(0)=\mathbf{a}$ and $[Z(0)]_{1}=a$.

From $(\ref{aaa})$, we know that
\begin{equation}
\begin{aligned}
\label{43}
&\Bigg|\sum_{i=1}^{t}\lambda_{1}^{t-i}(A(K,G))\left[P^{*}[
                I_{nN},
                \mathbf{0}^{T}
]^{T}(I_{N}\otimes
B)H(i)\right]_{1}\\
&~+\sum_{i=0}^{t}\lambda_{1}^{t-i}(A(K,G))g(i)\left[P^{*}[
                I_{nN},
                \mathbf{0}^{T}
]^{T}(I_{N}\otimes G)\Delta(i)\right]_{1}\Bigg|\cr
&~\leq\left(\frac{2W_{u}\|B\|\sqrt{mN}}{1-\varrho}+\frac{2L_{G}W\sqrt{nN}}{1-\varrho}\right)|\lambda_{1}(A(K,G))|^{t+1}\\
&~<\frac{a}{2}|\lambda_{1}(A(K,G))|^{t+1}.
\end{aligned}
\end{equation}
From $(\ref{z1})$, $(\ref{43})$ and noting that $H(0)=\mathbf{0}$, we have
\begin{equation*}
 \begin{aligned}
  \big|[Z(t+1)]_{1}\big|&\geq\left||\lambda_{1}(A(K,G))|^{t+1}a-\frac{a}{2}|\lambda_{1}(A(K,G))|^{t+1}\right|\\
  &=\frac{a}{2}|\lambda_{1}(A(K,G))|^{t+1}.
\end{aligned}
\end{equation*}
By the invertibility of $P$, we know that $[E^T(t), \delta^T(t)]^T$
does not vanish as $t\to\infty$. This is in contradiction with that the
dynamic network achieves inter-agent state observation and
cooperative stabilization. So, \textbf{A1)} and \textbf{A2)} hold. \qed
\vskip 0.2cm

\begin{remark}
{\rm Actually, the communication protocol parameter
$\gamma$ can represent the convergence speed of the cooperative
coordination (for both inter-agent state observation and cooperative
stabilization). The smaller $\gamma$ is, the faster the convergence
will be. The constant $\varrho$ is an upper bound of $\gamma$, so it is
a uniform upper bound of the convergence speed. Theorem \ref{dingli} shows that if $(A, B,
C, \mathscr{G})$  is locally cooperatable with a uniform exponential
convergence speed, then \textbf{A1)} and \textbf{A2)} hold.}
\end{remark}

\vskip 0.2cm

\begin{remark}
{\rm
Sundaram and Hadjicostis (\cite{Sundaram and Hadjicostis (2013)}) showed that a linear system is structurally controllable and observable over a finite field if the graph of the system satisfies certain properties and the size of the field is large enough. They also applied this result into the control of multi-agent systems over finite fields. Compared with \cite{Sundaram and Hadjicostis (2013)}, this note has the following differences.
(i) \cite{Sundaram and Hadjicostis (2013)} focused on the controllability and observability of linear systems over finite fields, and the closure property of the finite field plays an important role in getting their results. In this note we study the quantized coordination of linear multi-agent systems over real number field, so the closure and invertible properties can not be used. (ii) The system matrix $A$ of the linear system in \cite{Sundaram and Hadjicostis (2013)}  corresponds to the graph structure of the whole network, and the dynamics of each agent is actually in some integrator form. What is more, the elements of the system matrices $A$, $B$ and $C$ are restricted in finite fields. In this note, the affect of the graph topology is decided by the Laplacian matrix, and each agent has the general linear dynamics (see (\ref{1})), where the system matrices $A$, $B$ and $C$ are arbitrary real matrices.

As preliminary research, this note is concerned with inter-agent state observation and cooperative stabilization of multi-agent systems over digital networks. It is an interesting topic for further investigation that whether our results can be combined with the methodology of \cite{Sundaram and Hadjicostis (2013)} to study the controllability of multi-agent networks under quantized communication.}
\end{remark}

\vskip 0.2cm

At present, we still do not know whether \textbf{A1)} and \textbf{A2)} are necessary
conditions for $(A, B, C,\mathscr{G})$ to be locally cooperatable
w. r. t. $\mathscr{H}(1,+\infty)$ and $\mathscr{U}(+\infty)$. However,
we can show that if $(A,B,C,\mathscr{G})$ is globally cooperatable,
then \textbf{A1)} and \textbf{A2)} are necessary w. r. t. $\mathscr{H}(1,+\infty)$ and $\mathscr{U}(+\infty)$.

\vskip 0.2cm

\begin{theorem}\label{dingli123123123}
For $(A, B, C, \mathscr{G})$ and $L_{G}=+\infty$, $L_{K}=+\infty$
and $\varrho=1$, if there exist a communication protocol $H(\gamma,
\alpha, \alpha_{u}, L, L_u, G)\in\mathscr{H}(\varrho, L_{G})$ and a
control protocol $U(K)\in\mathscr{U}(L_{K})$, such that for any
$X(0)\in \mathbb{R}^{nN}$, $\hat{X}(0)\in \mathbb{R}^{nN}$ and
$\hat{U}(0)\in \mathbb{R}^{mN}$, the closed-loop system achieves
inter-agent state observation and cooperative stabilization under
$H$ and $U$, and $\sup_{t\geq0}\max_{1\leq j\leq
N}\|\Delta_{j}(t)\|_{\infty}$ $<\infty$ and $\sup_{t\geq0}\max_{1\leq
j\leq N}\|\Delta_{u,j}(t)\|_{\infty}<\infty$, then Assumptions
\textbf{A1)} and \textbf{A2)} hold.
\end{theorem}

\vskip 0.2cm

From the following theorems, we can see that the stabilizability of
$(A, B)$ is necessary for $(A, B, C, \mathscr{G})$ to achieve
cooperative stabilization no matter whether the inter-agent state
observation is required, and similarly, the detectability of $(A, C)$ is necessary for $(A, B, C, \mathscr{G})$ to achieve inter-agent state observation regardless of the cooperative stabilization.

\vskip 0.2cm

\begin{theorem}\label{ABnengwen}
For $(A, B, C, \mathscr{G})$, $L_{G}=+\infty$, $L_{K}=+\infty$ and
$\varrho=1$, suppose that for any given positive constants $C_{x}$,
$C_{\hat{x}}$ and $C_{\hat{u}}$, there exist a communication
protocol $H(\gamma, \alpha, \alpha_{u}, L, L_u,
G)\in\mathscr{H}(\varrho, L_{G})$ and a control protocol
$U(K)\in\mathscr{U}(L_{K})$, such that for any $X(0)\in
\mathscr{B}^{nN}_{C_{x}}$, $\hat{X}(0)\in
\mathscr{B}^{nN}_{C_{\hat{x}}}$ and
$\hat{U}(0)\in\mathscr{B}^{mN}_{C_{\hat{u}}}$, the closed-loop
system achieves cooperative stabilization under $H$ and $U$,  that
is, $\lim_{t\to\infty}(x_j(t)-x_i(t))=\textbf{0}$, $\forall\ i, j=1,2,...,N$.
Then $(A,B)$ is stabilizable.
\end{theorem}

\begin{theorem}\label{ACnengjian}
For $(A, B, C, \mathscr{G})$, $L_{G}=+\infty$, $L_{K}=+\infty$ and
$\varrho=1$, suppose that for any given positive constants $C_{x}$,
$C_{\hat{x}}$ and $C_{\hat{u}}$, there exist a communication
protocol $H(\gamma, \alpha, \alpha_{u}, L, L_u,
G)\in\mathscr{H}(\varrho, L_{G})$ and a control protocol
$U(K)\in\mathscr{U}(L_{K})$, such that for any $X(0)\in
\mathscr{B}^{nN}_{C_{x}}$, $\hat{X}(0)\in
\mathscr{B}^{nN}_{C_{\hat{x}}}$ and
$\hat{U}(0)\in\mathscr{B}^{mN}_{C_{\hat{u}}}$, the closed-loop
system achieves inter-agent under $H$ and $U$,
then $(A,C)$ is detectable.
\end{theorem}


\section{Numerical example}\label{IV}

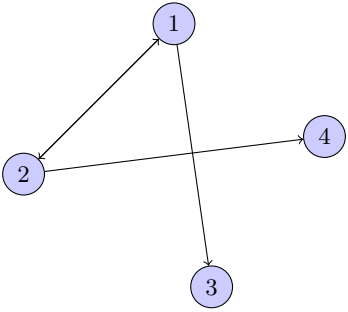
\begin{figure}[!tbh]
\begin{center}
\begin{tikzpicture}
\draw
   node at (0,0)[sum](1) {$1$}
   node at (-2,-2) [sum] (2) {$2$}
   node at (2, -1.5) [sum] (4) {$4$}
   node at (0.5,-3.5) [sum] (3) {$3$};
   \draw[->](1) -- node {} (2);
   \draw[->](2) -- node {} (1);
   \draw[->](1) -- node {} (3);
   \draw[->](2) -- node {} (4);
\end{tikzpicture}
\caption{The communication topology graph of the dynamic network.
}\label{tongxintu}
\end{center}
\end{figure}

\begin{figure}[!tbh]
\begin{center}
\includegraphics[width=7cm,height=6cm]{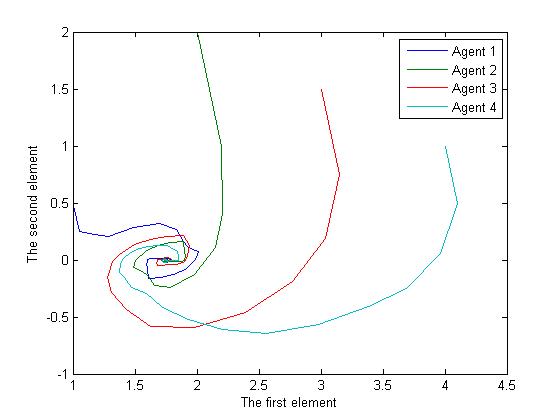}
\caption{The trajectories of agents' states.
}\label{fangzhen}
\end{center}
\end{figure}

\begin{figure}[!tbh]
\begin{center}
\includegraphics[width=7cm,height=6cm]{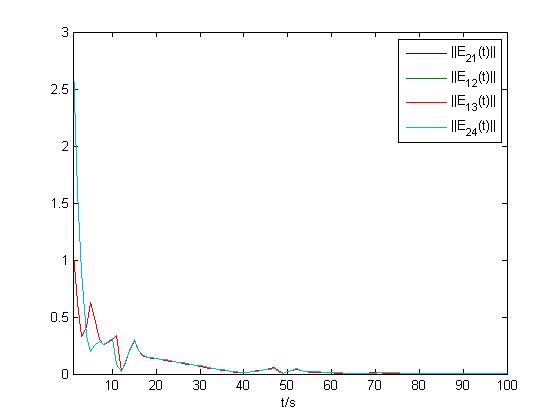}
\caption{The Euclidean norm of the state estimation errors.
}\label{error}
\end{center}
\end{figure}

Here, we consider a dynamic network with 4 agents. The state and measurement equations of each agent are given by
\begin{equation}
\left\{
    \begin{aligned}
     &x_{i}(t+1)=\left(
                 \begin{array}{cc}
                   1 & 0.1 \\
                   0 & 0.5 \\
                 \end{array}
               \right)x_{i}(t)+\left(
                                 \begin{array}{c}
                                   1 \\
                                   1 \\
                                 \end{array}
                             \right)u_{i}(t),\\
    & y_{i}(t)=\left(
                 \begin{array}{cc}
                   1 & 0 \\
                 \end{array}
               \right)x_{i}(t).
\end{aligned}
\right.
\end{equation}
The communication network is a directed 0-1 weight graph given by Figure \ref{tongxintu}.
We take the initial values of the agents randomly in the square: $[0,5]\times [0,5]$. We take $K=(0.2,0),G=(0.5,0)^{T}
$, $\gamma=0.95,\alpha=\alpha_{u}=1,L=L_{u}=20$.
Figure \ref{fangzhen} shows the evolution of agents' states.
Figure \ref{error} shows the Euclidean norm of the state estimation error associated with each digital channel.
From Figures \ref{fangzhen} and \ref{error}, it can be seen that the closed-loop system achieves both
cooperative stabilization and inter-agent state observation.

\section{Conclusion}\label{V}

In this note, we studied the inter-agent state observation and cooperative stabilization of discrete-time linear
multi-agent systems with unmeasurable states over bandwidth limited digital networks.
We proposed
a class of quantized-observer based communication
protocols and a class of Certainty Equivalence principle based control protocols.
We showed that
the simultaneous stabilizability condition and the detectability condition of agent dynamics
are sufficient for the existence of communication and control protocols to ensure
both the inter-agent state observation and cooperative stabilization. Furthermore, we proved that they are also necessary
for the local and global cooperatability in some sense. The theoretic results are also verified by numerical simulations.

As a preliminary research, we focus on the conditions on the dynamics of agents and the network structure
to ensure the existence of finite data rate inter-agent communication and control protocols.
An interesting
topic for future investigation is whether there is a lower bound, which is independent of the number of agents,
for the communication data rate required just as the small channel capacity theorems established
in \cite{Li Fu Xie and Zhang (2011)}, \cite{Li and Xie (2011)}, \cite{Liu Li and Xie (2011)} and \cite{Li and Xie (2012)}.
Note that the independency of the number of agents implies good scalability for large scale networks.
The problem is more challenging. Also,
Due to the time-delay, link failure or packet dropouts in networks, how to
design communication and control protocols for linear multi-agent systems to ensure both the cooperative stabilization and inter-agent state observation with finite data rate, communication delay and packet dropouts is an interesting and challenging problem.


\section*{Appendix}

\setcounter{lemma}{0}
\def\thelemma{A.\arabic{lemma}}
\def\thedefinition{A.\arabic{definition}}
\setcounter{equation}{0}
\def\theequation{A.\arabic{equation}}

\begin{lemma}\cite{You and Xie (2011)}
\label{you youxiang}
Assuming $A\in \mathbb{R}^{n\times n},B\in \mathbb{R}^{n\times 1}$, $\lambda_{2}$,
$\cdots$, $\lambda_{N}$, $i=2, \cdots, N$ are nonzero complex numbers. Assume that all the eigenvalues of $A$
are on or outside the unit circle of the complex plane. Then there exists
$K\in \mathbb{R}^{1\times n}$ such that $\rho(A-\lambda_{j}BK)<1$,\ $\forall$ $j=2, \cdots, N$ if and only if:

\qquad (a)\quad $(A,B)$ is controllable.

\qquad (b)\quad $\prod_{j}|\lambda_{j}(A)|<\frac{1}{\min_{\omega\in
\mathbb{R}}\max_{j\in\{2,\cdots,N\}}|1-\omega\lambda_{j}|}$
\end{lemma}

\vskip 0.2cm

\begin{lemma}\cite{Guo (1993)}
\label{GUO}
For any $A\in \mathbb{C}^{n\times n}$ and
$\varepsilon>0$, we have
\begin{equation*}
    \| A^{k}\|\leq M\eta^{k},\quad \forall~k\geq0,
\end{equation*}
where
$M=\sqrt{n}(1+(2/\varepsilon))^{n-1},~\eta=\rho(A)+\varepsilon\|
A\|$.
\end{lemma}

\vskip 0.2cm

\noindent
\textbf{Proof of Theorem \ref{youxiangyinli}}:
According to whether the state matrix $A$ is asymptotically stable, we prove the theorem for two cases, respectively.

\textbf{Case (i)}\quad $\rho(A)\geq1$.
For this case, since \textbf{A1)} holds, we can see that $0$ is a single eigenvalue of the
Lapalacian matrix $\mathcal{L}$.
Take $K\in\mathscr{B}^{m\times n}_{+\infty}$
such that $\rho(\bar{J}(K))<1$.
Take $G\in \mathscr{B}^{n\times p}_{+\infty}$ such that
$\rho(J(G))<1$.
Take $\varepsilon>0$ and $\bar{\varepsilon}_{1}>0$ such that $\eta=\rho(J(G))+\varepsilon\|
J(G)\|<1$ and
$\bar{\eta}_{1}=\rho(\bar{J}(K))+\bar{\varepsilon}_{1}\|
 \bar{J}(K)\|<1$.
Denote
$M=\sqrt{nN}(1+(2/\varepsilon))^{nN-1}$ and $\bar{M}_{1}=\sqrt{n(N-1)}(1+(2/\bar{\varepsilon}_{1}))^{n(N-1)-1}$.
Take
$\gamma\in(\max\{\eta,\bar{\eta}_{1}\},1)$ and $\alpha,\alpha_{u}\in(0,1]$. Let
$W(B,K)=diag(J_{2},\cdots,J_{N})\otimes BK$,
 and denote $R(G,\gamma,C_{x},C_{\hat{x}},\alpha)=\\\max\left\{\sqrt{nN}M(C_{x}+C_{\hat{x}}),\frac{\alpha\sqrt{pN}M\|
G\|}{2(\gamma-\eta)}\right\},$
\ban
&&L(K,G,\gamma,C_{x},C_{\hat{x}},C_{\hat{u}},\alpha,\alpha_{u})\cr
&=&\|C\|_{\infty}\Bigg(R(G,\gamma,C_{x},C_{\hat{x}},\alpha)+\frac{\alpha_{u}\sqrt{mN}M\| B\|}{2\gamma(\gamma-\eta)} \cr &&+\frac{1}{\gamma}M\|B\|(\sqrt{Nn}\|\mathcal{L}\otimes K\|C_{\hat{x}}+\sqrt{mN}C_{\hat{u}})\Bigg),
\ean
and
\ban
&&L_{u}(K,G,\gamma,C_{x},C_{\hat{x}},C_{\hat{u}},\alpha,\alpha_{u})\cr
&=&\max\Bigg\{2C_{x}\sqrt{nN}\| \mathcal{L}\otimes K-\mathcal{L}\otimes KA+\mathcal{L}^{2}\otimes KBK\|\cr
&&+\sqrt{nN}\|\mathcal{L}\otimes K(A-GC)-\mathcal{L}^{2}\otimes KBK-\mathcal{L}\otimes K\|\cr
&&\cdot(C_{x}+C_{\hat{x}})
+\frac{\alpha\sqrt{pN}\|\mathcal{L}\otimes KG\|}{2}+\|I_{mN}+\mathcal{L}\otimes
KB\|\cr
&&\cdot\Big(\sqrt{nN}\| \mathcal{L}\otimes K\|
C_{\hat{x}}+\sqrt{mN}C_{\hat{u}}\Big),
\|\Phi^{-1}\|\|\Phi\|\|\mathcal{L}\otimes K\cr
&&-\mathcal{L}\otimes KA+\mathcal{L}^{2}\otimes KBK\|\Gamma(K,G,\gamma,C_{x},C_{\hat{x}},C_{\hat{u}},\alpha,\alpha_{u})\cr
&&+\|\mathcal{L}\otimes K(A-GC)-\mathcal{L}^{2}\otimes
KBK-\mathcal{L}\otimes
K\|\cr
&&\cdot\bigg(R(G,\gamma,C_{x},C_{\hat{x}},\alpha)
+\frac{\alpha_{u}M\|B\|\sqrt{mN}
}{2\gamma(\gamma-\eta)}+\frac{1}{\gamma}M\|
B\|\cr
&&\cdot(C_{\hat{x}}\sqrt{Nn}\|\mathcal{L}\otimes K\|+C_{\hat{u}}\sqrt{mN})\bigg)\cr
    &&+\frac{\alpha_{u}\sqrt{mN}\| I_{mN}
    +\mathcal{L}\otimes KB\|}{2\gamma}+\frac{\alpha\sqrt{pN}\|\mathcal{L}\otimes
    KG\|}{2}\Bigg\},
\ean
where
\ban
&&\Gamma(K,G,\gamma,C_{x},C_{\hat{x}},C_{\hat{u}},\alpha,\alpha_{u})\cr
&=&\max\Bigg\{2C_{x}\bar{M}_{1}\sqrt{nN},\frac{\bar{M}_{1}\|
W(B,K)\|}{(\gamma-\bar{\eta}_{1})}\cr
&&\cdot\bigg(R(G,\gamma,C_{x},C_{\hat{x}},\alpha)
+\frac{\alpha_{u}M\|B\|\sqrt{mN}}{2\gamma(\gamma-\eta)}+\frac{1}{\gamma}M\|
B\|\cr
&&\cdot\left(\sqrt{Nn}\| \mathcal{L}\otimes K\|C_{\hat{x}}+\sqrt{mN}C_{\hat{u}}\right)\bigg)\Bigg\}.
\ean

Select the number of quantization levels
$L>\frac{1}{\alpha}L(K,G,\gamma,C_{x},C_{\hat{x}},C_{\hat{u}},\alpha,\alpha_{u})-\frac{1}{2}$
and
$L_{u}>\frac{1}{\alpha_{u}}L_{u}(K,G,\gamma,C_{x},
C_{\hat{x}},C_{\hat{u}},\alpha,\alpha_{u})$-$\frac{1}{2}$. Next we prove that with the parameters
$\gamma,\alpha,\alpha_{u},L,L_{u},G$ of the communication protocol
and the parameter $K$ of the control protocol selected as above, the
dynamic network will achieve the inter-agent state observation and
cooperative stabilization, and the quantization errors will be uniformly
bounded.

We first prove that by selecting the parameters $\gamma$, $L$, $L_u$, $\alpha$, $\alpha_{u}$, $G$ of the
communication protocol and the parameter $K$ of the control protocol as above, the quantizers $Q_{\alpha,L}(\cdot)$ and $Q_{\alpha_{u},L_{u}}(\cdot)$
will never be saturate. Denote $\tilde{\delta}(t)=(\Phi\otimes
I_{n})\delta(t)$ and $\tilde{E}(t)=(\Phi\otimes I_{n})E(t)$. Let
$\tilde{\delta}(t)=[(\tilde{\delta}_{1}^{T}(t)),(\tilde{\delta}_{2}^{T}(t))]^{T}$,
where $\tilde{\delta}_{1}(t)$ is the first $n$ elements of
$\tilde{\delta}(t)$. So $\tilde{\delta}_{1}(t)=(\pi^{T}\otimes
I_{n})\delta(t)=0$. Let
$\tilde{E}(t)=[(\tilde{E}_{1}^{T}(t)),(\tilde{E}_{2}^{T}(t))]^{T}$,
where $\tilde{E}_{1}(t)$ is the first $n$ elements of
$\tilde{E}(t)$. From $(\ref{FULU})$, the definition of $\bar{J}(K)$, $W(B,K)$, $J(G)$, and $H(t+1)=\gamma^t(F(t)-Q_{\alpha_{u},L_{u}}(F(t))=\gamma^t\Delta_u(t)$, we have
\begin{equation}\label{delta2}
    \begin{aligned}
    \tilde{\delta}_{2}(k+1)
    &=\bar{J}(K)\tilde{\delta}_{2}(k)+W(B,K)\tilde{E}_{2}(k)\\
    &=(\bar{J}(K))^{k+1}\tilde{\delta}_{2}(0)+\sum_{i=0}^{k}(\bar{J}(K))^{k-i}W(B,K)\\
    &~\cdot\tilde{E}_{2}(i),\ k=0, 1,...,
\end{aligned}
\end{equation}
and
\begin{equation}\label{E(t+1)1}
      \begin{aligned}
      E(k+1)&=(J(G))^{k+1}E(0)+\sum_{i=1}^{k}(J(G))^{k-i}(I_{N}\otimes
    B)\gamma^{i-1}\\
    &~\cdot\Delta_{u}(i)+\sum_{i=0}^{k}(J(G))^{k-i}(I_{N}\otimes
    G)\Delta(i)
    +(J(G))^{k}\\
    &~\cdot(I_{N}\otimes B)H(0),\ k=0, 1,...
\end{aligned}
\end{equation}
By Lemma \ref{GUO}, we know that $\|(J(G))^{i}\|\leq
M\eta^{i}$, $i=0,1,...$. Note that
$U(0)=-(\mathcal{L}\otimes K)\hat{X}(0)$,
$\|\hat{U}(0)\|_{\infty}\leq C_{\hat{u}}$ and
$H(0)=U(0)-\hat{U}(0)$, we have $\|H(0)\|\leq C_{\hat{x}}\sqrt{Nn}\|
\mathcal{L}\otimes K\|+\sqrt{mN}C_{\hat{u}}$.
Then from $(\ref{E(t+1)1})$, we have
\begin{equation}\label{E(k+1)fanshu}
    \begin{aligned}
    \| E(k+1)\|&\leq
    M\eta^{k+1}\sqrt{nN}(C_{x}+C_{\hat{x}})+\sum_{i=1}^{k}M\eta^{k-i}\|
    B\|\\
    &~\cdot\gamma^{i-1}
    \sqrt{mN}\|\Delta_{u}(i)\|_{\infty}
    +\sum_{i=0}^{k}M\eta^{k-i}\|
    G\|
    \gamma^{i}\\
    &~\cdot\sqrt{pN}\|\Delta(i)\|_{\infty}+M\eta^{k}\|
    B\|\big(C_{\hat{x}}\sqrt{Nn}\| \mathcal{L}\otimes K\|\\
    &~+C_{\hat{u}}\sqrt{mN}\big).
\end{aligned}
\end{equation}
At the initial time $k=0$, it is known that
\begin{equation*}
\begin{aligned}
    \|(I_{N}\otimes
    C)E(0)\|_{\infty}&\leq\| C\|_{\infty}\|
    E(0)\|_{\infty}\leq\|
    C\|_{\infty}(C_{x}+C_{\hat{x}}).
    \end{aligned}
\end{equation*}
Then noting that $\|
    C\|_{\infty}(C_{x}+C_{\hat{x}})\leq\|C\|_{\infty}\sqrt{nN}M(C_{x}+C_{\hat{x}})$$\leq R(G,\gamma,C_{x},C_{\hat{x}},\alpha)<\alpha L+\frac{\alpha}{2}$, we know that  $Q_{\alpha,L}(\cdot)$ is not saturate at the initial time $k=0$,
which means $\| \Delta(0)\|_{\infty}\leq\frac{\alpha}{2}$. It can be seen that that $\|\delta(0)\|_{\infty}\leq 2C_{x}$. Then from $(\ref{FULU})$, we get that
\begin{equation*}
\begin{aligned}
&\| F(0)\|_{\infty}
\leq2C_{x}\sqrt{Nn}\| \mathcal{L}\otimes K-\mathcal{L}\otimes KA+\mathcal{L}^{2}\otimes KBK\|\\
&~~~~~~+ \sqrt{Nn}\|
\mathcal{L}\otimes K(A-GC)-\mathcal{L}^{2}\otimes KBK-\mathcal{L}\otimes K\|\\
&~~~~~~\cdot(C_{x}+C_{\hat{x}})
+\frac{\alpha\sqrt{pN}\|\mathcal{L}\otimes
KG\|}{2}+\| I_{mN}+\mathcal{L}\otimes
KB\|\\
&~~~~~~\cdot\left(\sqrt{Nn}\| \mathcal{L}\otimes K\|
C_{\hat{x}}+\sqrt{mN}C_{\hat{u}}\right)
< \alpha_{u} L_{u}+\frac{\alpha_{u}}{2}.
\end{aligned}
\end{equation*}
Thus we know that $Q_{\alpha_{u},L_{u}}(\cdot)$ is
also not saturate at $k=0$, which means $\|\Delta_{u}(0)\|_{\infty}\leq\frac{\alpha_{u}}{2}$. Assume that $Q_{\alpha,L}(\cdot)$ and
$Q_{\alpha_{u},L_{u}}(\cdot)$ are not saturate at $k=0,1,...,t$, which implies,
$\max_{0\leq k\leq t}\|\Delta(k)\|_{\infty}\leq\frac{\alpha}{2}$ and
$\max_{0\leq k\leq t}\|\Delta_{u}(k)\|_{\infty}\leq\frac{\alpha_{u}}{2}$.
Now consider the time $k=t+1$. From (\ref{E(k+1)fanshu}), we have
\ban
&&\left\|\frac{(I_{N}\otimes C)E(t+1)}{\gamma^{t+1}}\right\|_{\infty}\leq\| C\|_{\infty}\frac{\| E(t+1)\|}{\gamma^{t+1}}\cr
&&\leq\|C\|_{\infty}\Bigg(\max\left\{\frac{\sqrt{Nn}M(C_{x}+C_{\hat{x}})}{\gamma^{t+1}},\frac{\alpha\sqrt{pN}M\|G\|}{2(\gamma-\eta)\gamma^{t+1}}\right\}\cr
&&~\cdot\gamma^{t+1}
+\frac{\alpha_{u}\sqrt{mN}M\|B\|}{2\gamma(\gamma-\eta)}\cdot\frac{\gamma^{t}-\eta^{t}}{\gamma^{t}}\cr
&&+\frac{M\eta^{t}\| B\|(\sqrt{Nn}\| \mathcal{L}\otimes K\|C_{\hat{x}}+\sqrt{mN}C_{\hat{u}})}{\gamma^{t+1}}\Bigg)\cr
&&\leq\| C\|_{\infty}\bigg(R(G,\gamma,C_{x},C_{\hat{x}},\alpha)+\frac{\alpha_{u}\sqrt{mN}M\|
B\| }{2\gamma(\gamma-\eta)}\cr
&&+\frac{1}{\gamma}M\| B\|\left(\sqrt{nN}\| \mathcal{L}\otimes K\|C_{\hat{x}}+\sqrt{mN}C_{\hat{u}}\right)\bigg)\cr
&&<\alpha L+\frac{\alpha}{2}.
\ean
So $Q_{\alpha,L}(\cdot)$ is not saturate at $k=t+1$. By
(\ref{E(k+1)fanshu}), and noting that $Q_{\alpha,L}(\cdot)$ and $Q_{\alpha_{u},L_{u}}(\cdot)$
are not saturate at $k=0,\cdots,t$, we have
\begin{equation}\label{ei}
\begin{aligned}
    \| E(k)\| & \leq \gamma^{k}\Bigg(R(G,\gamma,C_{x},C_{\hat{x}},\alpha)
+\frac{\alpha_{u}\sqrt{mN}M\|
B\| }{2\gamma(\gamma-\eta)}\\
&~+\frac{1}{\gamma}M\| B\|\left(\sqrt{Nn}\| \mathcal{L}\otimes K\|
C_{\hat{x}}+\sqrt{mN}C_{\hat{u}}\right)\Bigg),\\
&~~\quad 0\leq k\leq t+1.
\end{aligned}
\end{equation}
Next we will prove that $Q_{\alpha_{u},L_{u}}(\cdot)$ is not saturate at the time
$k=t+1$.

Since $\max_{0\leq k\leq
t}\|\Delta_{u}(k)\|_{\infty}\leq\frac{\alpha_{u}}{2}$, we have $\|
H(k+1)\|=\gamma^{k}\|\Delta_{u}(k)\|\leq
\gamma^{k}\sqrt{mN}\|\Delta_{u}(k)\|_{\infty}\leq\frac{\alpha_{u}\sqrt{mN}\gamma^{k}}{2}$, $0\leq k\leq t$.
Then from (\ref{delta2}), (\ref{ei}), Lemma \ref{GUO}, and note that $\tilde{\delta}_{1}(t)=\mathbf{0}$, we
have
\bna
\label{delta k fanshu}
\|\tilde{\delta}(k+1)\|&=&\|\tilde{\delta}_{2}(k+1)\| \cr
&\leq&\|(\bar{J}(K))^{k+1}\|\|\Phi\|\|\delta(0)\|+\sum_{i=0}^{k}\|(\bar{J}(K))^{k-i}\|\cr
&&\cdot\|W(B,K)\|\|\Phi\|\|E(i)\|\cr
&\leq&2\sqrt{Nn}\| \Phi\| C_{x}\bar{M}_{1}\bar{\eta}_{1}^{k+1}\cr
&&+\|\Phi \|\frac{\bar{M}_{1}\|
W(B,K)\|(\gamma^{k+1}-\bar{\eta}_{1}^{k+1})}{(\gamma-\bar{\eta}_{1})}\cr
&&\cdot\bigg(R(G,\gamma,C_{x},C_{\hat{x}},\alpha)+\frac{\alpha_{u}\sqrt{mN}M\| B\|
}{2\gamma(\gamma-\eta)}\cr
&&+\frac{1}{\gamma}M\|
B\|\left(\sqrt{Nn}\| \mathcal{L}\otimes K\|
C_{\hat{x}}+\sqrt{mN}C_{\hat{u}}\right)\bigg)\cr
&\leq& \gamma^{k+1}\|\Phi\|\max\Bigg\{2\sqrt{Nn}C_{x}\bar{M}_{1},\frac{\bar{M}_{1}\| W(B,K)\|}{(\gamma-\bar{\eta}_{1})}\cr
&&\cdot\bigg(R(G,\gamma,C_{x},C_{\hat{x}},\alpha)
+\frac{\alpha_{u}\sqrt{mN}M\|
B\| }{2\gamma(\gamma-\eta)}\cr
&&+\frac{1}{\gamma}M\| B\|\left(\sqrt{Nn}\|
\mathcal{L}\otimes K\|
C_{\hat{x}}+\sqrt{mN}C_{\hat{u}}\right)\bigg)\Bigg\}\cr
&=&\gamma^{k+1}\|\Phi\|\Gamma(K,G,\gamma,C_{x},C_{\hat{x}},C_{\hat{u}},\alpha,\alpha_{u}),\cr
&&0\leq k\leq t.
\ena
Thus, from $H(t+1)=\gamma^t\Delta_u(t)$, $(\ref{FULU})$,  $(\ref{ei})$ and $(\ref{delta k fanshu})$, we know that
\ban
&&\|F(t+1)\|_{\infty}
\leq\|\Phi^{-1}\otimes I_{n}\|\|\Phi\otimes I_{n}\|\|\mathcal{L}\otimes K \cr
&&~~~~~~~~~~~~~~~~~-\mathcal{L}\otimes KA
+\mathcal{L}^{2}\otimes KBK\|\cr
&&~~\cdot\Gamma(K,G,\gamma,C_{x},C_{\hat{x}},C_{\hat{u}},\alpha,\alpha_{u})
+\|\mathcal{L}\otimes K(A-GC)\cr
&&~~-\mathcal{L}^{2}\otimes KBK-\mathcal{L}\otimes K\|\bigg(R(G,\gamma,C_{x},C_{\hat{x}},\alpha) \cr
&&~~+\frac{\alpha_{u}\sqrt{mN}M\|B\|}{2\gamma(\gamma-\eta)}+\frac{1}{\gamma}M\|B\|\big(\sqrt{Nn}\| \mathcal{L}\otimes K\|C_{\hat{x}}\cr
&&~~+\sqrt{mN}C_{\hat{u}}\big)\bigg)
    +\frac{\alpha_{u}\sqrt{mN}\|I_{mN}+\mathcal{L}\otimes KB\|}{2\gamma}\cr
    &&~~+\frac{\alpha\sqrt{pN}\|\mathcal{L}\otimes KG\|}{2}\cr
    &&~~\leq L_{u}(K,G,\gamma,C_{x},C_{\hat{x}},C_{\hat{u}})<\alpha_{u}L_{u}+\frac{\alpha_{u}}{2}.
\ean
So $Q_{\alpha_{u},L_{u}}(\cdot)$ is not saturate at $k=t+1$. By induction,
$Q_{\alpha,L}(\cdot)$ and $Q_{\alpha_{u},L_{u}}(\cdot)$ are not saturate at any time. Thus,
under the selected communication protocol and control protocol, we have
$\sup_{t\geq0}\max_{1\leq j\leq N}\|
\Delta_{j}(t)\|_{\infty}\leq1/2$ and
$\sup_{t\geq0}\max_{1\leq j\leq
N}\|\Delta_{u,j}(t)\|_{\infty}\leq1/2$.

Now we prove that the dynamic network will achieve inter-agent state
observation and cooperative stabilization. Similar to
(\ref{ei}) and noting that
$\sup_{t\geq0}\max_{1\leq j\leq N}\|
\Delta_{j}(t)\|_{\infty}\leq \frac{1}{2}$ and\\
$\sup_{t\geq0}\max_{1\leq j\leq N}\|
\Delta_{u,j}(t)\|_{\infty}\leq\frac{1}{2}$, we have
\begin{equation}\label{e t wuqiong}
\begin{aligned}
    \| E(t)\| &\leq \gamma^{t}\Bigg(R(G,\gamma,C_{x},C_{\hat{x}},\alpha)+\frac{\alpha_{u}\sqrt{mN}M\|
B\| }{2\gamma(\gamma-\eta)}\\
&~+\frac{1}{\gamma}M\| B\|\left(\sqrt{Nn}\| \mathcal{L}\otimes K\|
C_{\hat{x}}+\sqrt{mN}C_{\hat{u}}\right)\Bigg),\\
&~\ t=0,1,...,
\end{aligned}
\end{equation}
which means
$\lim_{t\rightarrow\infty}E(t)=0$, that is, the dynamic network achieves
inter-agent state observation. Similar to (\ref{delta k fanshu}), by (\ref{delta2}) and (\ref{e t wuqiong}), we have
\begin{equation*}
\begin{aligned}
   \|\delta(t)\| & \leq \|\Phi^{-1}\otimes I_{n} \|\|\tilde{\delta}(t)\|=\|\Phi^{-1}\|\|\tilde{\delta}_{2}(t)\|\\
   &\leq\|\Phi^{-1}\|\|\Phi\|\Gamma(K,G,\gamma,C_{x},C_{\hat{x}},C_{\hat{u}},\alpha_{u})\gamma^{t},\\
   &~\ t=0,1,...,
\end{aligned}
\end{equation*}
which implies $\|\delta(t)\|_{\infty}\rightarrow0$, that is, the dynamic network achieves cooperative stabilization.

\textbf{Case (ii)}\quad $\rho(A)<1$.
Take $K=\mathbf{0}$ and $G=\mathbf{0}$, then we have $\bar{J}(K)=I_{N-1}\otimes
A,J(G)=I_{N}\otimes A$ and $W(B,K)=\mathbf{0}$. Take $\alpha$ and $\alpha_{u}\in
(0,1]$.
Denote
\begin{equation*}
\begin{aligned}
&L(\gamma,C_{x},C_{\hat{x}},C_{\hat{u}})=\|C\|_{\infty}\bigg(\sqrt{Nn}M(C_{x}+C_{\hat{x}})\\
&~~~~~~~~~+\frac{\alpha_{u}\sqrt{mN}M\|
B\|}{2\gamma(\gamma-\eta)}+\frac{1}{\gamma}M\|B\|\sqrt{mN}C_{\hat{u}}\bigg),
\end{aligned}
\end{equation*}
and
\ban
L_{u}(\gamma,C_{x},C_{\hat{x}},C_{\hat{u}})=\max\left\{\sqrt{mN}C_{\hat{u}},
\frac{\alpha_{u}\sqrt{mN}}{2\gamma}\right\}.
\ean
Select
$L>\frac{1}{\alpha}L(\gamma,C_{x},C_{\hat{x}},C_{\hat{u}})-\frac{1}{2}$
and $
L_{u}>\frac{1}{\alpha_{u}}L_{u}(\gamma,C_{x},C_{\hat{x}},C_{\hat{u}})-\frac{1}{2}$,
then similar to  \textbf{Case (i)}, one can prove that the
dynamic network will achieve inter-agent state observation and
cooperative stabilization, and the quantization errors are uniformly bounded
under the communication protocol and control protocol selected
above. \qed

\vskip 0.2cm

\noindent
\textbf{Proof of Theorem \ref{youkeyoutuilun}:}

\textbf{Case(i)}\quad $\rho(A)\geq1$.
Firstly, we use the reduction to absurdity to prove that
$\lambda_{j}(\mathcal{L})\neq0$, $j=2,...,N$. If not, there is an integer $k_{0}\in[2,N]$  such
that $\lambda_{k_{0}}(\mathcal{L})=0$.
It can be seen that $\inf_{\omega\in
R}\max_{j\in\{2,\cdots,N\}}|1-\omega\lambda_{j}(\mathcal{L})|\geq\inf_{\omega\in
R}\max_{j\in\{2,\cdots,N\}}|1-|\omega||\lambda_{j}(\mathcal{L})||$.
Denote $\max_{j\in\{2,\cdots,N\}}|\lambda_{j}(\mathcal{L})|$ by $p$.
Since $\lambda_{k_{0}}(\mathcal{L})=0$, we know that
$\min_{j\in\{2,\cdots,N\}}|\lambda_{j}(\mathcal{L})|=0$, thus, one get that
\begin{equation}\label{A52}
 \begin{aligned}
 &\max_{j\in\{2,\cdots,N\}}|1-|\omega||\lambda_{j}(\mathcal{L})||=\max\{|1-|\omega|\max_{j}|\lambda_{j}(\mathcal{L})||,\\
 &~~~~|1-|\omega|\min_{j}|\lambda_{j}(\mathcal{L})||\}\\
 &~~~~=\max\{1,|1-|\omega||p|\}=\left\{\begin{array}{c}
                                                                              -\omega p-1\quad \omega<-\frac{2}{p}, \\
                                                                              1\quad -\frac{2}{p}\leq\omega\leq \frac{2}{p},\\
                                                                              \omega
                                                                              p-1\quad\omega>\frac{2}{p}.
                                                                            \end{array} \right.
                                                                            \end{aligned}
\end{equation}
From $(\ref{A52})$, we know that $\inf_{\omega\in
R}\max_{j\in\{2,\cdots,N\}}|1-|\omega||\lambda_{j}(\mathcal{L})||=1$.
So $\inf_{\omega\in
R}\max_{j\in\{2,\cdots,N\}}|1-\omega\lambda_{j}(\mathcal{L})|
\geq\min_{\omega\in
R}\max_{j\in\{2,\cdots,N\}}|1-|\omega||\lambda_{j}(\mathcal{L})||=1$.
Then by A1$^{'}$), we know that
$\prod_{j}|\lambda_{j}^{u}(A)|<\frac{1}{\inf_{\omega\in
R}\max_{j\in\{2,\cdots,N\}}|1-\omega\lambda_{j}(\mathcal{L})|}\leq1$, which means $\rho(A)<1$. However,
$\rho(A)\geq1$ for case (1), so
$\lambda_{j}(\mathcal{L})\neq0,~j=2,\cdots,N$.

Next we prove that there exists $K\in\mathbb{R}^{1\times n}$ such that
$A-\lambda_{i}(\mathcal{L})BK,i=2,\cdots,N$ are all asymptotically stable. Denote the block diagonal
matrix which consists of the stable Jordan blocks of $A$ as
$A_{s}\in R^{n_{s}\times n_{s}},n_{s}\geq0$, and the block diagonal
matrix which consists of the other Jordan blocks of $A$ as $A_{u}\in
R^{n_{u}\times n_{u}},n_{u}\geq0$. For this case, since $\rho(A)\geq1$, we know that
$n_{u}>0$. Thus, there exists an invertible matrix $T$ such that
\begin{equation}\label{xin}
T^{-1}AT=\left(
                                                            \begin{array}{cc}
                                                              A_{s} & \mathbf{0} \\
                                                              \mathbf{0} & A_{u} \\
                                                            \end{array}
                                                          \right).
\end{equation}
 Let $T^{-1}B=\left(
                                                                                                   B_{1}^{T},                                                                            B_{2}^{T}
\right)^{T}$,
where $B_{1}\in\mathbb{R}^{n_{s}\times 1}$ and $B_{2}\in\mathbb{R}^{n_{u}\times 1}$. Now we prove that the matrix pair $(A_{u},B_{2})$
is controllable. If not, transform $(A_{u},B_{2})$ into its
controllable canonical, that is, there is an invertible matrix $R$ such that
\begin{equation}\label{57}
    R^{-1}A_{u}R=\left(\begin{array}{cc}
                         A_{u1} & A_{u2} \\
                         \mathbf{0} & A_{u3}
                       \end{array}
    \right),\quad R^{-1}B_{2}=\left(\begin{array}{c}
                                      B_{21} \\
                                      \mathbf{0}
                                    \end{array}
    \right),
\end{equation}
where $A_{u1}\in\mathbb{R}^{n_{u1}\times n_{u1}}$, $A_{u3}\in\mathbb{R}^{n_{u3}\times n_{u3}}$ and $A_{u3}$ is unstable. Then we know that
\begin{equation}\label{TR}
\begin{aligned}
    & A=T\left(\begin{array}{cc}
               I_{n_{s}} & \mathbf{0} \\
               \mathbf{0} & R
             \end{array}
    \right)\left(\begin{array}{ccc}
                   A_{s} & \mathbf{0} & \mathbf{0} \\
                   \mathbf{0} & A_{u1} & A_{u2} \\
                   \mathbf{0} & \mathbf{0} & A_{u3}
                 \end{array}
    \right)\left(\begin{array}{cc}
               I_{n_{s}} & \mathbf{0} \\
               \mathbf{0} & R^{-1}
             \end{array}
    \right)\\
    &~\cdot T^{-1},
     B=T\left(\begin{array}{cc}
               I_{n_{s}} & \mathbf{0} \\
               \mathbf{0} & R
             \end{array}
    \right)\Big(
                   B_{1}^{T},
                   (B_{21},
                   \mathbf{0})^{T}
    \Big)^{T}.
\end{aligned}
\end{equation}
For any given $K\in \mathbb{R}^{1\times n}$, let $KT\left(\begin{array}{cc}
               I_{n_{s}} & \mathbf{0} \\
               \mathbf{0} & R
             \end{array}
    \right)=(\hat{K}_{1},\hat{K}_{2},\hat{K}_{3})$, where $\hat{K}_{1}\in\mathbb{R}^{1\times n_{s}},\hat{K}_{2}\in\mathbb{R}^{1\times n_{u1}}$ and $\hat{K}_{3}\in\mathbb{R}^{1\times n_{u3}}$, we can see that
\bna
&&A+BK =T\left(\begin{array}{cc}
               I_{n_{s}} & \mathbf{0} \\
               \mathbf{0} & R
             \end{array}
    \right)\Big[\left(\begin{array}{ccc}
                   A_{s} & \mathbf{0} & \mathbf{0} \\
                   \mathbf{0} & A_{u1} & A_{u2} \\
                   \mathbf{0} & \mathbf{0} & A_{u3}
                 \end{array} \right) \cr
    &&+\left(\begin{array}{c}
                    B_{1} \\
                    B_{21} \\
                    \mathbf{0}
                  \end{array}
    \right)(\hat{K}_{1},\hat{K}_{2},\hat{K}_{3})\Big]\left(\begin{array}{cc}
               I_{n_{s}} & \mathbf{0} \\
               \mathbf{0} & R^{-1}
             \end{array}
    \right)T^{-1} \cr
    && =T\left(\begin{array}{cc}
               I_{n_{s}} & \mathbf{0} \\
               \mathbf{0} & R
             \end{array}
    \right)\cr
    &&\cdot\left(\begin{array}{ccc}
                   A_{s}+B_{1}\hat{K}_{1} & B_{1}\hat{K}_{2} & B_{1}\hat{K}_{3} \\
                   B_{21}\hat{K}_{1} & A_{u1}+B_{21}\hat{K}_{2} & A_{u2}+B_{21}\hat{K}_{3} \\
                   \mathbf{0} & \mathbf{0} & A_{u3}
                 \end{array}
    \right)\cr
    &&\cdot\left(\begin{array}{cc}
               I_{n_{s}} & \mathbf{0} \\
               \mathbf{0} & R^{-1}
             \end{array}
    \right)T^{-1}.
\ena
Since $A_{u3}$ is unstable, we know that there is no matrix $K$ such that
$A+BK$ is stable, which is in contradiction with that $(A,B)$ is stabilizable. So
we know that $(A_{u},B_{2})$ controllable.

Since
$\lambda_{i}(\mathcal{L})\neq0,i=2,\cdots,N$, from Lemma \ref{you
youxiang}, we know that there exist a $\tilde{K}\in R^{1\times n_{u}}$
such that
$\rho(A_{u}-\lambda_{i}(\mathcal{L})B_{2}\tilde{K})<1,i=2,\cdots,N$.
Take $\bar{K}=[\textbf{0}^{T},\tilde{K}]$ and take $K=\bar{K}T^{-1}$,
since
\begin{equation*}
    T^{-1}AT=\left(
                                                            \begin{array}{cc}
                                                              A_{s} & \mathbf{0} \\
                                                              \mathbf{0} & A_{u} \\
                                                            \end{array}
                                                          \right), T^{-1}B=\left(\begin{array}{c}
                                                                                                   B_{1} \\
                                                                                                   B_{2}
                                                                                                 \end{array}\right),
\end{equation*}
we know that
\begin{equation*}
\begin{aligned}
A-\lambda_{i}(\mathcal{L})BK&=T\left(
                                                            \begin{array}{cc}
                                                              A_{s} & -\lambda_{i}(\mathcal{L})B_{1}\tilde{K} \\
                                                              \mathbf{0} & A_{u}-\lambda_{i}(\mathcal{L})B_{2}\tilde{K} \\
                                                            \end{array}
                                                          \right)T^{-1},\\
                                                          &~i=2,\cdots,N.
\end{aligned}
\end{equation*}
 So $\rho(A-\lambda_{i}(\mathcal{L})BK)<1$, $i=2, \cdots, N$.

So \textbf{A1$^{'}$)} suffices for \textbf{A1)}.

\textbf{Case (ii)}\quad $\rho(A)<1$.
For this case, take $K=\mathbf{0}^{T}$, then \textbf{A1)} holds.
\qed

\vskip 0.2cm

\noindent
\textbf{Proof of Theorem \ref{dingli123123123}}: Denote $\sup_{t\geq0}\max_{1\leq j\leq
N}\|\Delta_{j}(t)\|_{\infty}$ by $W$ and $\sup_{t\geq0}$$\max_{1\leq j\leq N}$ $\|\Delta_{u,j}(t)\|_{\infty}$ by $W_{u}$. Noting that here, different from
Theorem \ref{dingli}, $W$ and $W_{u}$ may depend on the parameters $\gamma$, $\alpha,\alpha_{u}$, $L,L_{u}$, $G$ and $K$.
Select a constant $a$ satisfying
 \begin{equation}\label{aaaa}
 a>\frac{4\|
B\|\sqrt{mN}W_{u}}{|\lambda_{1}(A(K,G))-\gamma|}+\frac{4\|
G\| W\sqrt{nN}}{|\lambda_{1}(A(K,G))-\gamma|}.
 \end{equation}

Take $X(0)=(\Phi^{-1}\otimes I_{n})[\mathbf{0}^T,\mathbf{a}^TP_{2}^T]^T$ where $\mathbf{a}=a\mathbf{1}\in\mathbb{R}^{n(2N-1)}$ and $\mathbf{0}\in\mathbb{R}^{n}$. Take
$\hat{X}(0)=X(0)-P_{1}\mathbf{a},\hat{U}(0)=-(\mathcal{L}\otimes
K)\hat{X}(0)$, thus $E(0)=X(0)-\hat{X}(0)=X(0)-(X(0)-P_{1}\mathbf{a})=P_{1}\mathbf{a}$,
    and $H(0)=U(0)-\hat{U}(0)=-(\mathcal{L}\otimes K)\hat{X}(0)+(\mathcal{L}\otimes K)\hat{X}(0)=\mathbf{0}$. Thus
$Z(0)=\mathbf{a}$ and $[Z(0)]_{1}=a$.
Then similar to the proof of Theorem\ref{dingli}, we have the conclusion.\qed

\vskip 0.2cm

\noindent
\textbf{Proof of Theorem \ref{ABnengwen}:}
We use the reduction to absurdity to prove this theorem. Suppose that $(A,B)$ is unstabilizable, then there exists an invertible matrix $T_{1}$, such that $T_1^{-1}AT_1=\left(\begin{array}{cc}
                                                              A_{s_1} &  A_{12}\\
                                                              \mathbf{0} & A_{u4} \\
                                                            \end{array}
                                                          \right)$ and $T_1^{-1}B=\left(B_{3}^T, \mathbf{0}^T\right)^T$,
where  $A_{u4}\in\mathbb{R}^{n_{u4}\times n_{u4}}$ is unstable. Here $n_{u4}$ is a positive integer. Take $C_{x}>\sqrt{n}\|\Phi^{-1}\|\|T_1\|$. Take $C_{\hat{x}}>1$ and $C_{\hat{u}}>1$. Next we prove that for any given communication protocol in (\ref{rongxutongxinxieyiji}) and
control protocol in (\ref{rongxukongzhi}), there exist
$X(0)\in \mathscr{B}^{nN}_{C_{x}}$, $\hat{X}(0)\in\mathscr{B}^{nN}_{C_{\hat{x}}}$ and
$\hat{U}(0)\in\mathscr{B}^{mN}_{C_{\hat{u}}}$, such that the dynamic
network can not achieve cooperative stabilization, which leads to the
contradiction.
Denote the first $n$ elements of $\tilde{E}_{2}(t)$ and $\tilde{\delta}_{2}(t)$ by $\tilde{E}_{21}(t)$ and $\tilde{\delta}_{21}(t)$ where $\tilde{E}_{2}(t)$ and
$\tilde{\delta}_{2}(t)$ are defined in the proof of Theorem \ref{youxiangyinli}. From (\ref{FULU}), we have
\begin{equation}\label{63}
\tilde{\delta}_{21}(t+1)=(A-\lambda_{2}(\mathcal{L})BK)\tilde{\delta}_{21}(t)+\lambda_{2}(\mathcal{L})BK\tilde{E}_{21}(t).
\end{equation} Denote
    $\hat{\delta}_{21}(t)=T_1^{-1}\tilde{\delta}_{21}(t)$, and let  $KT_1=(\hat{K}_{3},\hat{K}_{4})$ where $\hat{K}_{3}\in\mathbb{R}^{m\times (n-n_{u4})},\hat{K}_{4}\in\mathbb{R}^{m\times n_{u4}}$.
Thus, from $(\ref{63})$,  we have
\begin{equation}\label{65}
\begin{aligned}
 \hat{\delta}_{21}(t+1)&=
 \left(\begin{array}{cc}
                   A_{s1}-\lambda_{2}(\mathcal{L})B_{3}\hat{K}_{3}  & \times \\
                   \mathbf{0}  & A_{u4}
                 \end{array}
    \right) \hat{\delta}_{21}(t)\\
    &~+\left(\begin{array}{cc}
                   \lambda_{2}(\mathcal{L})B_{3}\hat{K}_{3}&  \lambda_{2}(\mathcal{L})B_{3}\hat{K}_{4} \\
                   \mathbf{0} & \mathbf{0}
                 \end{array}
    \right)T_1^{-1}\tilde{E}_{21}(t).
\end{aligned}
\end{equation}
 Denote the last
$n_{u4}$ elements of $\hat{\delta}_{21}(t+1)$ by
$\hat{\delta}_{21_{n_{u4}}}(t+1)$, thus from $(\ref{65})$, we have
$\hat{\delta}_{21_{n_{u4}}}(t+1)=A_{u4}\hat{\delta}_{21_{n_{u4}}}(t)$.
Take $X(0)=(\Phi^{-1}\otimes I_{n})\left(
\mathbf{0}^{T},
\left[T_1\mathbf{1}\right]^{T}
    \right)^{T}$, then
    $\|X(0)\|_{\infty}\leq
\sqrt{n}\|\Phi^{-1}\|
\| T_1\|<C_{x}$. By the definition of $\delta(t)$, and noting that $\pi^{T}$ is the first row of $\Phi$, we have
$\delta(0)=(\Phi^{-1}\otimes I_{n})\left(\mathbf{0}^{T},\left[T_1\mathbf{1}\right]^{T}\right)^{T}$.
Thus, $\hat{\delta}_{21}(0)=\mathbf{1}_{n}$ and $\hat{\delta}_{21_{n_{u4}}}(0)=\mathbf{1}_{n_{u4}}$.
Take $\hat{X}(0)=\mathbf{1}_{nN},\hat{U}(0)=\mathbf{1}_{mN}$, then we have
$\|\hat{X}(0)\|_{\infty}<C_{\hat{x}}$ and $\|
\hat{U}(0)\|_{\infty}<C_{\hat{u}}$ . Since $\hat{\delta}_{21_{n_{u4}}}(0)\neq0$,
$\delta(t)$ does not vanish, which draws the contradiction.
\qed

\vskip 0.2cm

\textbf{Proof of Theorem \ref{ACnengjian}:} We use the reduction to absurdity to prove this theorem.  If $(A,C)$ was not detectable, then there would exist $x_{0}\in\mathbb{R}^{n}$, such that $CA^{l}x_{0}=\textbf{0}$, $l=0,1,2,...,$, and $A^{t}x_{0}$ does not go to zero as $t\to\infty$. Take $C_{x}>\|x_{0}\|$, $C_{\hat{x}}>0$ and $C_{\hat{u}}>0$. Next we will prove that for any given communication protocol $H\in\mathscr{H}(1, +\infty)$ and control protocol $U\in\mathscr{U}(+\infty)$, there exist $X(0)\in \mathscr{B}^{nN}_{C_{x}}$, $\hat{X}(0)\in\mathscr{B}^{nN}_{C_{\hat{x}}}$ and
$\hat{U}(0)\in\mathscr{B}^{mN}_{C_{\hat{u}}}$, such that the dynamic
network can not achieve inter-agent state observation, which leads to the
contradiction. Take $x_{1}(0)=x_{0}$ and $x_{j}=\textbf{0}$, $j=2,\cdots,N$. Then $X(0)\in\mathscr{B}^{nN}_{C_{x}}$.  By $Cx_{0}=\textbf{0}$, we have $y_j(0)=\textbf{0}$, $j=1,2,...,N$. Take $\hat{X}(0)=\textbf{0}$ and  $\hat{U}(0)=\textbf{0}$, so $\hat{X}(0)\in\mathscr{B}^{nN}_{C_{\hat{x}}}$ and
$\hat{U}(0)\in\mathscr{B}^{mN}_{C_{\hat{u}}}$. By (\ref{rongxukongzhi}), we know that $U(0)=\textbf{0}$. By (\ref{bianma}), (\ref{jiema}), noting that $y_j(0)=\textbf{0}$, $j=1,2,...,N$, we know that
$s_{j}(1)=\textbf{0}$, $j=1,2,...,N$, which together with
$\hat{X}(0)=\textbf{0}$ and $\hat{U}(0)=\textbf{0}$ lead to $\hat{X}(1)=\textbf{0}$. Then by (\ref{bianma}), (\ref{jiema}) and  (\ref{rongxukongzhi}), it follows that $U(1)=\textbf{0}$, and $\hat{U}(1)=\textbf{0}$. Then by (\ref{1}) and $CAx_{0}=\textbf{0}$, we have $y_j(1)=0$, $j=1,2,...,N$. Suppose that up to time $t$, $t=2,3,...$, $U(k)=\hat{U}(k)=\textbf{0}$, and $\hat{X}(k)=\textbf{0}$, $k=0,1,...,t-1$. Then By (1), we have $x_{1}(t-1)=A^{t-1}x_0$, $x_{j}(t-1)=\textbf{0}$, $j=2,3,...,N$. Noting that $CA^{t-1}x_0=\textbf{0}$, it follows that $y_{j}(t-1)=0$, $j=1,2,...,N$. And by (\ref{bianma}), (\ref{jiema}) and (\ref{rongxukongzhi}), we know that $\hat{X}(t)=\textbf{0}$ and $U(t)=\hat{U}(t)=\textbf{0}$. Then by mathematical induction, we have $\hat{X}(t)\equiv\textbf{0}$ and $U(t)\equiv\textbf{0}$,
which together with (\ref{1}) gives $x_{1}(t)=A^{t}x_{0}$, and $x_{2}(t)=\cdots=x_{N}(t)\equiv\textbf{0}$. Noting that $A^{t}x_{0}$ does not go to zero as $t\to\infty$, but $\hat{X}(t)\equiv\textbf{0}$, it follows that $E(t)=X(t)-\hat{X}(t)$ does not go to zero as $t\to\infty$, which leads to the contradiction. \qed

\end{CJK*}


\small
\begin{thebibliography}{99}
\setlength{\itemsep}{0pt} \setlength{\parskip}{0pt}


%
%
%

\bibitem{Qu (2009)}
 Z. H. Qu,
{\em Cooperative control of dynamical systems: applications to autonomous
vehicles}.
 London: Springer-Verlag, 2009.

\bibitem{Kashyap Basar and Srikant (2007)}
A. Kashyap, T. Basar, and R. Srikant,
 ``Quantized consensus,"
 {\em Automatica},
 Vol. 43, No. 7, pp. 1192-1203, 2007.

\bibitem{Frasca Carli Fagnani and Zampieri (2009)}
P. Frasca, R. Carli, F. Fagnani, and S. Zampieri,
 ``Average
consensus on networks with quantized communication,''
{\em International Journal of Robust and Nonlinear
Control}, Vol. 19, No. 16, pp. 1787-1816, 2009.

\bibitem{Carli Bullo and Zampieri (2010)}
R. Carli, F. Bullo, and S. Zampieri,
``Quantized average consensus
via dynamic coding/decoding schemes,''
 {\em International Journal of Robust and Nonlinear
Control}, Vol. 20, No. 2, pp. 156-175, 2010.

\bibitem{Li Fu Xie and Zhang (2011)}
T. Li, M. Fu, L. Xie, and J. F. Zhang,
``Distributed consensus with
limited communication data rate,''
{\em IEEE Trans.
on Automatic Control}, Vol. 56, No. 2, pp. 279-292, 2011.

\bibitem{Li and Xie (2011)}
T. Li and L. Xie,
``Distributed consensus over digital networks with
limited bandwidth and time-varying topologies,''
{\em Automatica},
 Vol. 47, No. 9, pp. 2006-2015, 2011.




\bibitem{Li Liu Wang and Yin (2014)}
D. Li, Q. Liu, X. Wang and Z. Yin,
``Quantized consensus over directed networks with switching topologies,''
{\em Systems and Control Letters}, Vol. 65, No. 3, pp. 13-22, 2014.

\bibitem{Liu Li and Xie (2011)}
S. Liu, T. Li and L. Xie,
``Distributed consensus for multiagent systems with communication delays and limited data rate,''
 {\em SIAM Journal on Control and Optimization}, Vol. 49, No. 6, pp. 2239-2262, 2011.

\bibitem{Olshevsky (2014)}
A. Olshevsky,
``Consensus with ternary messages,''
{\em SIAM Journal on Control and Optimization}, Vol. 52, No. 2, pp. 987-1009, 2014.

\bibitem{Frasca (2012)}
P. Frasca,
``Continuous-time quantized consensus: Convergence of Krasovskii solutions,''
{\em Systems and Control Letters},
Vol. 61, No. 2, pp. 273-278, 2012.


\bibitem{Pasqualetti Borra and Bullo (2014)}
F. Pasqualetti, D. Borra and F. Bullo,
``Consensus networks over finite fields,''
{\em Automatica}, Vol. 50, No. 2, pp. 349-358, 2014.

%

%





%

\bibitem{Li and Xie (2012)}
T. Li and L. Xie,
``Distributed coordination of multi-agent systems
with quantized observer based encoding-decoding,''
 {\em IEEE Trans. on Automatic Control},
Vol. 57, No. 12, pp. 3023-3037, 2012.

\bibitem{Zhang and Tian (2009)}
 Y. Zhang and Y. P. Tian,
 ``Consentability and protocol design of multi-agent systems with stochastic
switching topology,'' {\em Automatica},
Vol. 45, No. 5, pp. 1195-1201, 2009.

\bibitem{Ma and Zhang (2010)}
 C. Q. Ma and J. F. Zhang,
``Necessary and sufficient conditions for
consensusability of linear multi-agent systems,''
{\em IEEE Trans.
on Automatic Control},
Vol. 55, No. 5, pp. 1263-1268, 2010.

\bibitem{You and Xie (2011)}
K. You and L. Xie,
``Network topology and communication data rate
for consensusability of discrete-time multi-agent systems,''
{\em IEEE Transactions
on Automatic Control},
Vol. 56, No. 10, pp. 2262-2275, 2011.

\bibitem{Gu Marinovici and Lewis (2012)}
G. X. Gu, L. Marinovici, and F. L. Lewis,
``Consensusability of Discrete-Time
Dynamic Multiagent Systems,''
 {\em IEEE Transactions on Automatic Control}, Vol. 57, No. 8, pp. 2085-2089, 2012.


\bibitem{Olfati-Saber and Murray (2004)}
R. Olfati-Saber and R. M. Murray,
 ``Consensus problem in networks of
agents with switching topology and time-delays,''
 {\em IEEE Trans.
on Automatic Control},
 Vol. 49, No. 9, pp. 1520-1533, 2004.

\bibitem{LWCV2005}
G. Lafferriere, A. Williams, J. Caughman and J. J. P. Veerman, ``Decentralized
control of vehicle formations,'{\em ' Systems and Control Letters}, Vol. 54, No. 9,
pp. 899-910, 2005.


\bibitem{Ren and Beard (2008)}
W. Ren and R. W. Beard,
 {\em Distributed Consensus in Multi-Vehicle
Cooperative Control}.
London: Springer-Verlag, 2008.
\bibitem{Sundaram and Hadjicostis (2013)}
S. Sundaram and C. N. Hadjicostis,
``Structural controllability and observability of linear systems over finite fields with applications to multi-agent systems,''
{\em IEEE Trans.
on Automatic Control}, Vol. 58, No. 1, pp. 60-73, 2013.

\bibitem{Guo (1993)}
 L. Guo, {\em Time-varying Stochastic System---stability, estimation and
control}, Changchun: Jilin since and tecnology press, 1993.





\end{thebibliography}
\end{document}